%% file: main.tex
\newif\ifdraft
\draftfalse
\ifdraft
    \documentclass{sig-alternate}
\else
    \documentclass{sig-alternate}
\fi

\usepackage{balance}
\usepackage{amsmath,amssymb,amsfonts}
\usepackage{graphicx}
\usepackage{textcomp}
\usepackage{pifont}
\usepackage{booktabs}
\usepackage{glossaries} 
\usepackage{setspace}
\usepackage{listings}
\usepackage[linesnumbered]{algorithm2e}
\usepackage{siunitx} 
\usepackage{subcaption}
\usepackage{tikz}
\usepackage{xcolor, color, soul}

\usepackage{interval} 
\usepackage{duckuments} 
\usepackage{algorithmic} 
\usepackage{mathptmx} 
\usepackage{fancyhdr}
\usepackage[normalem]{ulem}
\usepackage[hyphens]{url}
\usepackage[sort,compress]{cite}
\usepackage[final]{microtype}
\usepackage[keeplastbox]{flushend}
\usepackage{multirow}
\usepackage{footmisc}
\usepackage{filecontents}
\usepackage{authblk}

\usepackage[bookmarks=true,breaklinks=true,letterpaper=true,colorlinks,citecolor=blue,linkcolor=blue,urlcolor=blue]{hyperref}

\input{macros}

\input{glossary}

\newif\ifarxiv
\arxivtrue

\ifarxiv
\usepackage{calc}
\fancypagestyle{firstpage}
{
    \fancyhead{}
  
  \pagenumbering{arabic}
  \fancyfoot[C]{\large\thepage}
}
\fi

\title{\X{}: Simultaneous Many-Row Activation \\for Reliable and High-Performance Computing \\in Off-the-Shelf DRAM Chips}

\author{Ismail Emir Yuksel}
\author{Yahya Can Tugrul}
\author{F. Nisa Bostanci}
\author{Abdullah Giray Yaglikci}
\author{Ataberk Olgun}
\author{Geraldo F. Oliveira}
\author{Melina Soysal}
\author{Haocong Luo} 
\author{Juan Gomez Luna}
\author{Mohammad Sadrosadati}
\author{Onur Mutlu}
\affil{ETH Zurich}
\date{}                     

\begin{document}
\counterwithout{lstlisting}{subsection}

\maketitle
\pagestyle{plain}

\input{sections/00_abstract}
\glsresetall
\input{sections/01_introduction}

\glsresetall
\input{sections/02_background}
\input{sections/03_motivation}
\input{sections/04_sim-act}
\input{sections/05_mechanism}

\input{sections/06_use-cases}

\input{sections/07_discussion}

\input{sections/08_related-work}

\input{sections/09_conclusion}


\bibliographystyle{IEEEtranS}
\bibliography{refs}

\input{sections/appendix}

\end{document}

%% file: macros.tex
\newcommand{\XLong}[0]{Long Mechanism Name}
\newcommand{\X}[0]{PULSAR}

\definecolor{gfored}{rgb}{0.580, 0.050, 0.211}
\definecolor{ao}{rgb}{0.007, 0.520, 0.867}
\definecolor{moegi}{rgb}{0.357, 0.537, 0.188}
\definecolor{jl}{rgb}{1.0, 0.2, 0.8}
\definecolor{brown(web)}{rgb}{0.65, 0.16, 0.16}
\definecolor{bisque}{rgb}{1.0, 0.89, 0.77}
\definecolor{nbs}{rgb}{0.88, 0.07, 0.37}
\definecolor{yt}{rgb}{0.58, 0.44, 0.86}
\definecolor{iy}{rgb}{0.0, 0.36, 0.05}

\newcommand{\dingZero}{\circledtest{0}}
\newcommand{\dingOne}{\circledtest{1}}
\newcommand{\dingTwo}{\circledtest{2}}
\newcommand{\dingThree}{\circledtest{3}}
\newcommand{\dingFour}{\circledtest{4}}

\newcommand{\one}{1)}
\newcommand{\two}{2)}
\newcommand{\three}{3)}
\newcommand{\four}{4)}

\newcommand{\param}[1]{\textcolor{black}{#1}} 
\newcommand{\xxx}[1]{\param{XXX}} 

\newcommand{\ignore}[1]{}

\ifdraft
    \usepackage[colorinlistoftodos,prependcaption,textsize=tiny]{todonotes}
    \usepackage{color,soul}

    \newcommand{\agy}[1]{\textcolor{gfored}{#1}}
    \newcommand{\agycomment}[1]{\todo[size=\scriptsize, linecolor=orange, bordercolor=orange, backgroundcolor=white]{\textcolor{gfored}{\textbf{@gy:} #1}}}
    \renewcommand{\agycomment}[1]{{\textcolor{gfored}{\textbf{\hl{[@gy:}} \hl{#1}\textbf{]}}}}

    \newcommand{\mscomment}[1]{\todo[size=\scriptsize, linecolor=orange, bordercolor=orange, backgroundcolor=white]{\textcolor{red}{\textbf{@MS:} #1}}}
    \renewcommand{\mscomment}[1]{{\textcolor{red}{\textbf{\hl{[@MS:}} \hl{#1}\textbf{]}}}}

    \newcommand{\atb}[1]{\textcolor{ao}{#1}}

    \newcommand{\atbcomment}[1]{\todo[size=\scriptsize, linecolor=orange, bordercolor=orange, backgroundcolor=white]{\textcolor{ao}{\textbf{@atb:} #1}}}
    \renewcommand{\atbcomment}[1]{{\textcolor{ao}{\textbf{\hl{[@atb:}} \hl{#1}\textbf{]}}}}

    \newcommand{\yct}[1]{\textcolor{yt}{#1}}
    \newcommand{\yctcomment}[1]{\todo[size=\scriptsize, linecolor=orange, bordercolor=orange, backgroundcolor=white]{\textcolor{yt}{\textbf{@yct:} #1}}}
    \renewcommand{\yctcomment}[1]{{\textcolor{yt}{\textbf{\hl{[@yct:}} \hl{#1}\textbf{]}}}}

    \newcommand{\gfcomment}[1]{\todo[size=\scriptsize, linecolor=orange, bordercolor=orange, backgroundcolor=white]{\textcolor{blue}{\textbf{@gf:} #1}}}
    
    \newcommand{\nb}[1]{\textcolor{nbs}{#1}}
    \newcommand{\nbcomment}[1]{\todo[size=\scriptsize, linecolor=orange, bordercolor=orange, backgroundcolor=white]{\textcolor{nbs}{\textbf{@nb:} #1}}}
    \renewcommand{\nbcomment}[1]{{\textcolor{nbs}{\textbf{\hl{[@nb:}} \hl{#1}\textbf{]}}}}

    \newcommand{\hluo}[1]{\textcolor{moegi}{#1}}
    \newcommand{\hluocomment}[1]{\todo[size=\scriptsize, linecolor=orange, bordercolor=orange, backgroundcolor=white]{\textcolor{moegi}{\textbf{@hluo:} #1}}}

    \newcommand{\iey}[1]{\textcolor{iy}{#1}}
    \newcommand{\ieyinline}[1]{{\color{iy}{\textbf{\hl{[@iey:}} \textit{\hl{#1}}\textbf{]}}}}
    \newcommand{\ieycomment}[1]{\todo[size=\scriptsize, linecolor=orange, bordercolor=orange, backgroundcolor=white]{\textcolor{iy}{\textbf{@iey:} #1}}}
    \renewcommand{\ieycomment}[1]{{\color{iy}{\textbf{\hl{[@iey:}} \textit{\hl{#1}}\textbf{]}}}}

    \newcommand{\om}[1]{\textcolor{teal}{#1}}
    \newcommand{\omcomment}[1]{\todo[size=\scriptsize, linecolor=orange, bordercolor=orange, backgroundcolor=white]{\textcolor{teal}{\textbf{@ste:} #1}}}
    \newcommand{\ominline}[1]{{\textcolor{teal}{\textbf{[@ste:} #1\textbf{]}}}}

\else

    \newcommand{\agy}[1]{{#1}}
    \newcommand{\agycomment}[1]{}
    \newcommand{\agyinline}[1]{}

    \newcommand{\mscomment}[1]{}
    
    \newcommand{\atb}[1]{{#1}}
    \newcommand{\atbcomment}[1]{}
    
    \newcommand{\yct}[1]{{#1}}
    \newcommand{\yctcomment}[1]{}

    \newcommand{\gfcomment}[1]{}

    \newcommand{\nb}[1]{{#1}}
    \newcommand{\nbcomment}[1]{}

    \newcommand{\hluo}[1]{{#1}}
    \newcommand{\hluocomment}[1]{}

    \newcommand{\iey}[1]{#1}
    \newcommand{\ieycomment}[1]{}
    \newcommand{\ieyinline}[1]{}
    
    \newcommand{\om}[1]{#1}
    \newcommand{\omcomment}[1]{}
    \newcommand{\ominline}[1]{}

\fi

\newcommand*\nCHIPS{120}
\newcommand*\nMODULES{21}

\newcounter{obs}
\newcounter{tkw}
\newcommand{\cosDRAMPumAllCitations}[0]{\cite{olgun2021quactrng, gao2019computedram, gao2022frac, kim2019drange, kim2018dram}}

\newcommand{\PUMAllCitations}[0]{\cite{aga2017compute, besta2021sisa,chi2016prime,deng2018dracc,eckert2018neural,flashcosmos,fujiki2019duality,gao2019computedram,hajinazar2020simdram,he2020sparse,imani2019floatpim,li2016pinatubo,li2017drisa,oliveira2022accelerating,seshadri.arxiv16,seshadri2013rowclone,seshadri2015fast,seshadri2016processing,seshadri2017ambit,seshadri2017simple,seshadri2018rowclone,seshadri2019dram,shafiee2016isaac,song2017pipelayer,song2018graphr,xin2020elp2im, gao2022frac}}

\newcommand{\dataMovementProblemsCitations}[0]{~\cite{
mutlu2013memory,
mutlu2015research,
dean2013tail,
kanev_isca2015,
ferdman2012clearing,
wang2014bigdatabench,
mutlu2019enabling,
mutlu2019processing,
kim2012case,
wulf1995hitting,
ghose.sigmetrics20,
ahn2015scalable,
PEI,
hsieh2016transparent,
wang2020figaro}}
\newcommand{\figref}[1]{Fig.~\ref{#1}}

\newcommand{\secref}[1]{§\ref{#1}}

\newcommand{\maj}[1]{\texttt{MAJ{#1}}}

\newcommand*\circledtest[1]{\tikz[baseline=(char.base)]{
            \node[shape=circle,fill,inner sep=0.3pt] (char) {\textcolor{white}{#1}};}}

%% file: glossary.tex
\newcommand{\tras}[0]{t_{RAS}}
\newcommand{\trp}[0]{t_{RP}}
\newcommand{\trc}[0]{t_{RC}}
\newcommand{\trefi}[0]{t_{REFI}}
\newcommand{\trefw}[0]{t_{REFW}}
\newcommand{\trfc}[0]{t_{RFC}}
\newcommand{\trrd}[0]{t_{RRD}}
\newcommand{\tfaw}[0]{t_{FAW}}

\newcommand{\nrg}[0]{$N_{RG}$}

\newcommand{\act}[0]{\texttt{ACT}}
\newcommand{\pre}[0]{\texttt{PRE}}
\newcommand{\refresh}[0]{REF}
\newcommand{\wri}[0]{\texttt{WR}}
\newcommand{\rd}[0]{\texttt{RD}}

\newcommand{\bralow}[0]{\lowercase{Multiple (up to 32) Row Activation}}

\newcommand{\apaLong}[0]{\act{} $\rightarrow$ \pre{} $\rightarrow$ \act{}}
\newcommand{\apa}[0]{\texttt{APA}}

\newcommand{\rfirst}[0]{$R_{F}$}
\newcommand{\rsecond}[0]{$R_{S}$}

\newcommand{\vdd}[0]{\texttt{VDD}}
\newcommand{\gnd}[0]{\texttt{GND}}
\newcommand{\vddh}[0]{\texttt{VDD/2}}

\newcommand{\apaEx}[0]{{\act{} \rfirst{} $\rightarrow$ \pre{} $\rightarrow$ \act{} \rsecond{}}}
\newcommand{\apaex}[0]{{\act{} 0 $\rightarrow$ \pre{} $\rightarrow$ \act{} 7}}

\newcommand{\pum}[0]{\texttt{PuM}}

\usepackage[shortcuts]{extdash}
\newcommand{\mrc}[0]{Multi\-/RowInit}
\newacronym{iqr}{$IQR$}{inter-quartile range}
\newacronym{act}{\act{}}{activate}
\newacronym{pre}{\pre{}}{precharge}
\newacronym{ref}{\refresh{}}{refresh}
\newacronym{wr}{\wri{}}{write}
\newacronym{rd}{\rd{}}{read}
\newacronym{pum}{\pum{}}{Processing-using-Memory}
\newacronym{apa}{\apa{}}{\act{} $\rightarrow$ \pre{} $\rightarrow$ \act{}}
\newacronym{mech}{\emph{\X{}}}{\XLong}
\newacronym{jedec}{JEDEC}{Joint Electron Device Engineering Council}

\newacronym{trefw}{$\trefw$}{refresh window}
\newacronym{tras}{$\tras$}{the latency of sensing the row's data and fully restoring a DRAM cell's charge}
\newacronym{trp}{$\trp$}{the latency of de-asserting a wordline and precharging the bitlines to \vddh{}}
\newacronym{trc}{$\trc$}{the minimum time needed between two consecutive row activations targeting the same bank}
\newacronym{trefi}{$\trefi$}{refresh interval}
\newacronym{trfc}{$\trfc$}{refresh latency}
\newacronym{trrd}{$\trrd$}{the minimum time needed between two consecutive row activations targeting the same rank}

\newacronym{puf}{PUF}{physical unclonable function}
\newacronym{trn}{TRN}{true random number}
\newacronym{tfaw}{$\tfaw{}$}{four activation window}

%% file: sections/00_abstract.tex
\begin{abstract}
\agy{Data movement between the processor and the main memory is a first-order obstacle against improving performance and energy efficiency in modern systems.}
\agy{To address this obstacle, \gls{pum} is a promising approach where bulk-bitwise operations are performed leveraging intrinsic analog properties within the DRAM array and massive parallelism across DRAM columns.}
\agy{Unfortunately, 1)~modern off-the-shelf DRAM chips do not officially support \gls{pum} operations and 2)~existing techniques of performing \gls{pum} operations on off-the-shelf DRAM chips suffer from two key limitations. First, these techniques} have low success rates\agy{, i.e., only a small fraction of DRAM columns can correctly execute \pum{} operations, because} they operate beyond manufacturer-recommended \agy{timing constraints,} \agy{causing} these operations to be highly susceptible to noise and process variation. \agy{Second,} these techniques have limited compute primitives, preventing them from fully leveraging parallelism across DRAM columns and thus hindering their performance benefits.

\agy{We propose \X{}, a new technique to enable high-success-rate and high-performance \gls{pum} operations in off-the-shelf DRAM chips.}
\X{} leverages \agy{our new} observation that a carefully-crafted sequence of DRAM commands simultaneously activates up to 32 DRAM rows.
\agy{\X{} overcomes the limitations of existing techniques by 1)~replicating the input data to improve the success rate and 2)~enabling new} bulk bitwise operations (e.g., many-input majority, \emph{\mrc{}}, and \emph{Bulk-Write}) to improve the performance. 

Our analysis on \nCHIPS{} off-the-shelf DDR4 chips from two major manufacturers shows that \X{} achieves $24.18\%$ higher success rate and $121\%$ higher performance over seven arithmetic-logic operations compared to FracDRAM, a state-of-the-art off-the-shelf DRAM-based \pum{} technique.
\end{abstract}

%% file: sections/01_introduction.tex
\section{Introduction}
\label{sec:introduction}

\agy{Data movement between the processor and the main memory is a first-order obstacle against improving performance and energy efficiency in modern systems~\cite{mutlu2019processing, boroumand2018google}.}
\agy{Many prior works~\PUMAllCitations{} propose} \gls{pum} \om{techniques} to alleviate data movement bottlenecks\agy{, where} computation is performed directly within memory arrays (e.g., DRAM) by leveraging the intrinsic analog operating properties of the memory device. \pum{} significantly \agy{reduces} data movement, \agy{thereby lowering both} \iey{energy consumption \agy{and} execution time \agy{(i.e., improving system performance).}}
\agy{\gls{pum} can be enabled in modern systems via 1)~}various modifications to DRAM \om{chips}~\cite{seshadri2017ambit,seshadri2015fast,flashcosmos,seshadri2019dram,seshadri2016processing,seshadri2017simple,seshadri.arxiv16,li2016pinatubo,xin2020elp2im,angizi2019redram} \agy{or 2)~violating the timing constraints without the need of any modifications to \om{DRAM chips}~\cosDRAMPumAllCitations{}}.

\agy{Prior work proposes \om{\pum{} techniques} that experimentally demonstrates that three sets of \gls{pum} operations can be executed in \om{unmodified} off-the-shelf DRAM chips:}
1)~bitwise \agy{logic} and arithmetic operations based on three-input majority function\atb{s}, i.e., \maj{3}~\cite{gao2019computedram, gao2022frac},
2)~\atb{bulk-data copy at DRAM row granularity~\cite{gao2019computedram,seshadri2013rowclone,olgun2022pidram} (\iey{called}, RowClone)}, and
3)~\agy{generating} security primitives (e.g., \atb{in-DRAM} true random number generation, physical unclonable functions~\cite{kim2019drange,kim2018dram,olgun2021quactrng,olgun2022pidram}). \atb{\agy{Unfortunately}, these operations suffer from two key \iey{problems} that significantly limit their applicability.}

\noindent
\textbf{Success Rate.} \om{\maj{3} operation is based on a multiple-row activation that connects \agy{a bitline to} multiple cells by simultaneously activating multiple DRAM rows.} \yct{We define the success rate} the percentage of bitlines that reliably and correctly perform \maj{3} operation. \om{Unfortunately}, \maj{3} operation\atb{s} in modern off-the-shelf DRAM chips have low success rate. \iey{This is because the \om{multiple-row activation} \one{} is \emph{not} officially supported by the DRAM manufacturers as it requires operating beyond manufacturer-recommended timing constraints, and \two{} can result in a smaller deviation on the bitline voltage than the reliable sensing margin due to noise and manufacturing process variation. These negative factors lead to preventing all bitlines from reliably and successfully performing \om{a} \maj{3} operation. }
 Prior work attempts to improve the success rate of \maj{3} in old-generation DRAM chips (i.e., DDR3)~\cite{gao2022frac}. However, the current off-the-shelf DRAM chips (i.e., DDR4) still suffer from low success rates \om{(e.g., 78.85\% on average across DDR4 chips we test)} in \maj{3} operations (\secref{subsubsec:motiv_success}). Consequently, \maj{3} in modern DRAM chips have poor reliability and frequently produce incorrect results.

\noindent
\atb{\textbf{Performance.}}
\om{\pum{} techniques~\cite{gao2019computedram,gao2022frac}} in unmodified off-the-shelf DRAM chips are limited in functionality, which significantly hinders their performance. 
\atb{Although these techniques can} perform basic operations such as two-operand bitwise AND/OR (e.g., \texttt{A}~\textbullet~\texttt{B}) and RowClone (\texttt{COPY A}~$\rightarrow$~\texttt{B})~\cite{gao2019computedram,olgun2022pidram,gao2022frac}, many modern applications \om{would benefit from executing} (e.g., data analytics~\cite{seshadri2017ambit,jun2015bluedbm,torabzadehkashi2019catalina}, databases~\cite{wu2005fastbit, wu1998encoded, guz2014real}, and graph processing~\cite{li2016pinatubo,besta2021sisa,hajinazar2020simdram}) more complex operations, such as  \iey{many-input (i.e., more than two)} bitwise AND/OR operations (e.g., \texttt{A}~\textbullet~\texttt{B}~\textbullet~\texttt{C}) and many row initialization (e.g., \texttt{COPY A}~$\rightarrow$~\texttt{[B, C]}). \agy{Due to limited functionality, prior} \atb{works  \emph{sequentially} execute the basic \pum{} operations to perform complex \pum{} operations.}
\atb{However, sequentially executing basic operations} leads to \atb{high} latency and \atb{low} throughput.

In this paper, we propose \X{}\footnote{We name our technique as \X{}, a \underline{P}\underline{u}M Technique that \underline{L}everages \underline{S}imultaneous \underline{A}ctivation of Many \underline{R}ows.}, \atb{a new} \pum{} technique that improves the \agy{success rate} and performance of \pum{} operations in \om{unmodified} off-the-shelf DRAM chips. We \agy{experimentally} demonstrate \agy{using} \nCHIPS{} off-the-shelf DDR4 DRAM chips from two major DRAM manufacturers \agy{that a carefully crafted sequence of DRAM commands simultaneously activates \om{many rows (i.e., 32)}.} \X{} leverages this \om{new} observation and demonstrates a proof-of-concept where off-the-shelf DRAM chips can be used to execute \pum{} operations with \om{much higher} \agy{success rate} and performance \atb{than \agy{the} state-of-the-art~\cite{gao2022frac}}.
\X{} overcomes the \om{two key problems of} the existing techniques~\cite{gao2022frac,gao2019computedram} by \one{} replicating the input data across different DRAM rows to improve the success rate and \two{} enabling \emph{new} \pum{} operations (e.g., \mrc{}, many-input charge-sharing operations, and Bulk-Write) to provide significant performance \om{improvements}.
\noindent
\textbf{Input Replication.} \X{} \agy{replicates (i.e.,} stores multiple copies of\agy{)} each majority operation's input on all simultaneously activated rows.
\agy{During \iey{multiple} row activation, these multiple} copies contribute to charge sharing and thus increase the net deviation in bitline voltage. For example, performing a \maj{3} operation \om{by simultaneously activating six rows that contain two copies} of each input results in \param{44.06\%} higher net deviation in bitline voltage than \om{activating three rows that store only one} copy of each input (\secref{subsec:inp_repl}). Larger deviation in bitline voltage
\om{greatly reduces the} effects of electrical noise and process variation on the results of \maj{3} operations. We \agy{present the first characterization}
\agy{of} the success rate of \maj{3} operations \om{in DDR4} using \nCHIPS{} off-the-shelf \agy{DRAM} chips. Our results show that \X{} executes \maj{3} operations with a \param{97.91}\% success rate, which is \param{24.18}\% higher than \om{that of} the state-of-the-art technique~\cite{gao2022frac}.

\noindent
\textbf{New PuM Primitives.}
Activating N rows simultaneously \om{(where N is up to 32) in off-the-shelf DRAM chips}, \om{enables} more complex operations. \X{} introduces new \pum{} primitives that perform bulk data operations on \agy{multiple (up to \param{N})} operands with a single simultaneous activation: \emph{\mrc{}}, many-input charge-sharing operations, and \emph{Bulk-Write}.
The \mrc{} primitive allows for the copying of one row into N rows simultaneously. Many input charge-sharing operations enable majority operations with up to \param{N} inputs (e.g., \maj{5} and \maj{7}). The Bulk-Write operation enables writing to N rows with \om{only one write command}. These primitives increase the throughput and reduce the latency of \agy{two} \pum{} operations\agy{:} 1) majority-based computation and 2) cold-boot-attack defense.

\noindent
\textbf{Majority-based Computation.}  To our knowledge, for the first time, we demonstrate a proof-of-concept that off-the-shelf DRAM chips can execute \maj{M} \yct{operations }(i.e., \maj{3}+ operations) with high reliability. We study the throughput and \agy{the} latency of majority-based computations in off-the-shelf DRAM chips using arithmetic and \iey{logic} operations. Our results show that \X{} improves performance by \om{121\%} on average compared to the \agy{state-of-the-art} technique~\cite{gao2022frac}.

\noindent
\textbf{Cold-Boot-Attack Defense.} We propose content destruction for cold boot attacks that leverage the new \pum{} primitives that \X{} introduces. Our results show that \X{} \om{speeds up} content destruction \om{in off-the-shelf DRAM chips} by \param{7.75}$\times$ compared to the FracDRAM~\cite{gao2022frac}-based content destruction technique.

\om{This paper makes} the following key contributions:
\begin{itemize}
    \item \om{We demonstrate, through an extensive experimental characterization of \yct{\nCHIPS{} }modern DRAM chips \yct{from two major manufacturers} that modern DRAM chips can simultaneously activate up to 32 DRAM rows.}
    \item We introduce \agy{\X{}}, a new \agy{\pum{}} technique \om{that leverages simultaneous activation of up to 32 rows.} \X{} improves the success rate and performance of \pum{} operations in off-the-shelf DRAM chips. \agy{\X{}} demonstrates a proof-of-concept that off-the-shelf DRAM chips are able to execute \maj{3} operations with a \param{97.91}\% success rate, which is \param{24.18}\% higher than the state-of-the-art~\cite{gao2022frac}.
    \item To our knowledge, for the first time, \agy{\X{}} demonstrates more than three-inputs \maj{} operations with a very high success rate \om{(73.93\% for \maj{5} on average across the DRAM modules that we test)} and core primitive\iey{s} called \mrc{} and \iey{Bulk-Write} that significantly reduces the latency of many row initialization.
    \item We show that \agy{\X{}} significantly improves the performance of \om{seven arithmetic and logic operations} over the state-of-the-art mechanism~\cite{gao2022frac} by \param{2.21}$\times$ and significantly reduce the latency of content destruction for cold boot attack in off-the-shelf-DRAM by \param{7.55}$\times$.
\end{itemize}

%% file: sections/02_background.tex
\section{Background}
\label{sec:background}
This section briefly details DRAM organization, operation, timings, and \pum{} operations in off-the-shelf DRAM chips.
\subsection{DRAM Organization}
\begin{figure}[ht]
\centering
\includegraphics[width=0.85\linewidth]{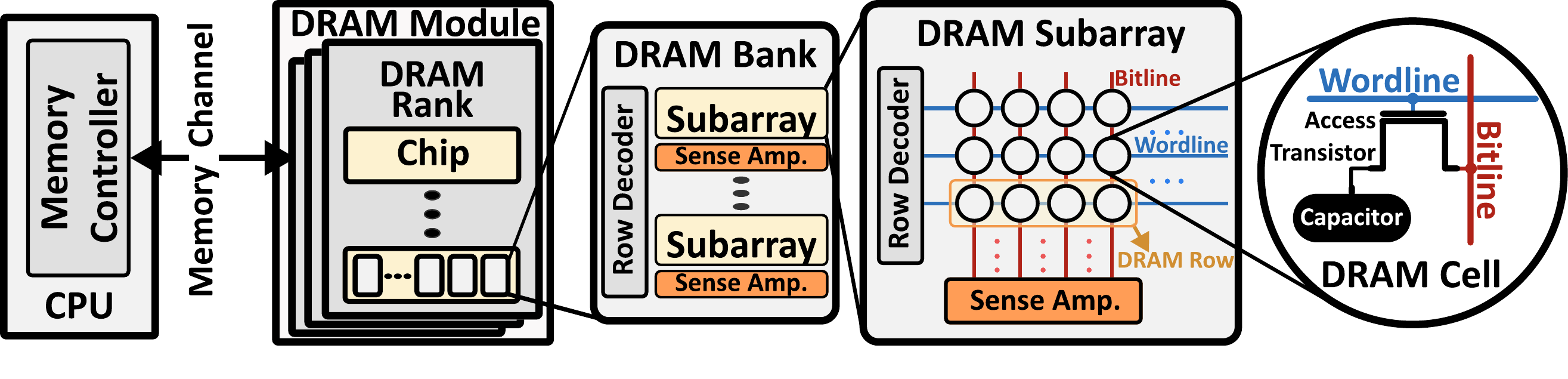}
\caption{DRAM Organization.}
\label{fig:dram_organization}
\end{figure}

\agy{\figref{fig:dram_organization} shows the organization of DRAM-based memory systems. A memory channel connects the processor (CPU) to \yct{a DRAM module where a module consists of multiple DRAM ranks. A rank is formed by} a set of DRAM chips operated in lockstep. } 
\agy{A} DRAM \agy{chip} \agy{has} multiple \textit{DRAM banks} \agy{each of which is composed of many DRAM subarrays.}
Within a subarray, DRAM cells form a two-dimensional structure interconnected over \textit{bitlines} and \textit{wordlines}. \yct{The row decoder in a subarray decodes the row address and drives the wordline out of many.} A row of DRAM cells on the same wordline is referred to as a DRAM row. \yct{The DRAM cells in the same column are connected to the sense amplifier via a bitline.}
A DRAM cell stores the binary data value \agy{in the form of electrical charge on a capacitor~\yct{(\vdd{} or 0 V)} and this data is accessed through an access transistor, which is driven by the wordline to conduct the cell capacitor to the bitline.}

\subsection{DRAM Operation and Timing}
\noindent
\textbf{\agy{Operation.}}
\agy{Data stored in a DRAM array is internally accessed in a DRAM row granularity.}
\yct{In the closed state, all wordlines are de-asserted and all bitlines are precharged to \vddh{} in a bank. To access a row, the data needs to be fetched to the sense amplifier. To do so, the memory controller issues an \act{} command to assert the wordline and enable the sense amplifier. When the wordline is asserted, the cell capacitor connects to the bitline and shares its charge causing a small voltage deviation on the bitline voltage. After, the sense amplifier is enabled to sense and amplify the small voltage deviation towards \vdd{} or 0 V, depending on the cell data. Once the data is fetched to the sense amplifiers and the cell's data is restored, the memory controller may issue \wri{}/\rd{} commands to write to/read from the row. To access another row, the bank needs to be in the closed (i.e., precharged) state. To do so, the memory controller issues a \pre{} command to disable sense amplifiers, de-assert the wordline, and precharging the bitlines to \vddh{}. Once the bank is precharged, the memory controller can access another row.}

\noindent
\textbf{Timing.}
\agy{To} ensure correct operation, the memory controller must obey the DRAM timing parameters specified in the \agy{DRAM interface standards (e.g., DDR4~\cite{jedec2017ddr4}) by \gls{jedec}}.
\yct{We describe the most relevant timing constraints in the scope of this paper. The memory controller must wait for \gls{tras} before issuing a \pre{} command after an \act{} command. To open another row, the memory controller must wait for \gls{trp} before issuing another \act{} command.}

\subsection{PuM Operations \agy{in Off-the-Shelf DRAM}}
PuM architectures allow computations to be performed within the memory array in contrast to the traditional architectures, where data has to be constantly transferred between memory and processor units. Off-the-shelf DRAM chips are not officially designed to support \pum{} operations (i.e., operations that are performed inside of a memory).
\agy{Although DRAM manufacturers or JEDEC do \emph{not} officially support \pum{} operations, the design of off-the-shelf DRAM chips does \emph{not} fully prevent users from activating mltiple at once by violating \gls{tras} and \gls{trp} timing constraints~\cite{gao2019computedram, olgun2021quactrng, gao2022frac, yauglikcci2022hira}. By doing so, two fundamental \pum{} operations can be performed in off-the-shelf DRAM chips: \one{} MAJ3 and \two{} RowClone}.

\noindent
\textbf{MAJ3.} 
 Prior work introduces the idea of multiple row activation (i.e., activating more than one row simultaneously) in off-the-shelf DRAM \yct{that enables charge sharing operation between multiple cells and leads to }
\maj{3} \yct{operation} across activated rows (i.e., row groups). State-of-the-art mechanism~\cite{gao2022frac} performs four-row activation to enable \maj{3} operations in off-the-shelf DRAM chips. 
However, with an even number of operands, the \maj{3} cannot be performed due to the equilibrium state (i.e., an equal number of ones and zeros). To address this issue, they propose Frac operation~\cite{gao2022frac}. Frac operation can charge any row to \texttt{VDD/2}, putting the row into a neutral state during multiple row activation. As a result, they enable \maj{3} by activating four rows at once.

\noindent
\textbf{RowClone~\cite{seshadri2013rowclone}.} Prior work~\cite{gao2019computedram} enables consecutive activation of two DRAM rows to copy data in DRAM, RowClone, in off-the-shelf DRAM chips. RowClone enables data movement within DRAM in a DRAM row granularity without incurring the energy and execution time costs of transferring data between the DRAM and the computing units.

\subsection{Majority-based Computation}
\label{back:maj}
\begin{figure}[ht]
\centering
\includegraphics[width=\linewidth]{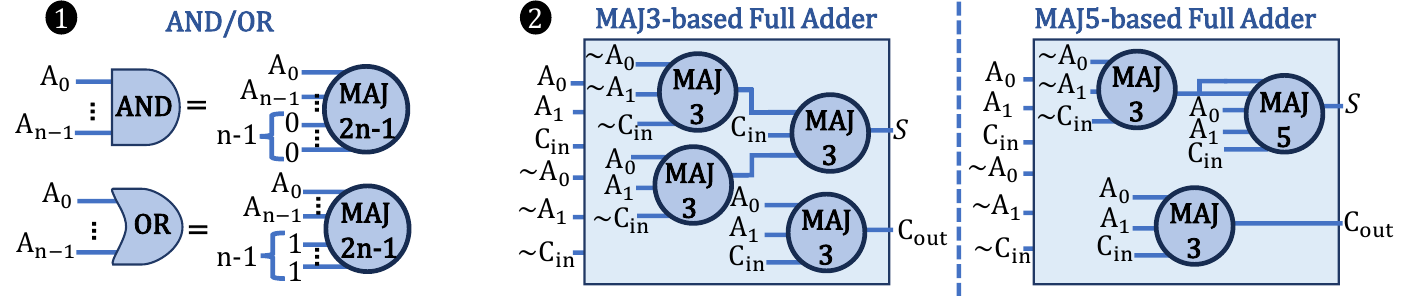}
\caption{Example of AND/OR and full-adder implementation using only MAJ gates with regular and negated data.}
\label{fig:maj_ops}
\end{figure}

Majority gates can be used to implement 1)~logic operations such as AND/OR~\cite{hajinazar2020simdram,ali2019memory,seshadri.arxiv16,seshadri2017ambit,seshadri2015fast,seshadri2019dram,gao2019computedram}) and XOR operations~\cite{alkaldy2014novel}, and 2)~full adders~\cite{ali2019memory,gao2019computedram,hajinazar2020simdram}. These operations are then used as basic building blocks for the target in-DRAM computation (e.g., addition, multiplication)~\cite{ali2019memory,angizi2019graphide,li2016pinatubo,gao2019computedram}. However, MAJ gates 
cannot implement a NOT operation. Therefore, it is not possible to implement building blocks that require the NOT gate (e.g., XOR operation and full adder) with only MAJ gates. 
Prior work~\cite{gao2019computedram} overcomes 
\nb{this limitation }
in off-the-shelf DRAM chips by storing both the regular and the negated version of a value. The presence of both regular and negated data allows us to perform any arbitrary function as we can implement functionally-complete logic gates (e.g., NAND, NOR).~\figref{fig:maj_ops} shows an example of AND/OR (\dingOne{}) and full-adder design (\dingTwo{}) using only majority gates with regular and negated inputs.

\noindent\textbf{Vertical Data Layout.}~Supporting bit-shift operations is essential for implementing complex computations, such as addition (e.g., carry propagation).~Prior works~\cite{hajinazar2020simdram,ali2019memory,gao2019computedram,seshadri2017ambit} provide this support by employing a vertical layout for the data in DRAM, such that all bits of an operand are placed in a single DRAM column (i.e., in a single bitline).~\agy{Doing so}
eliminates the need for adding extra logic in DRAM to implement shifting and applies bulk bitwise operations to entire rows of DRAM, generating results from bitlines in parallel.

%% file: sections/03_motivation.tex
\section{Motivation}
\label{sec:motivation}
Modern computing systems require moving data back and forth between computing units (e.g., CPU, GPU) and off-chip main memory to perform computation on the data~\cite{mutlu2019processing, boroumand2018google}. Unfortunately, this data movement is a major bottleneck that consumes a large fraction of execution time and energy~\dataMovementProblemsCitations{}. To address this problem, \gls{pum} emerges as a promising execution paradigm to alleviate the data movement bottleneck in the modern and emerging applications~\cite{seshadri2017ambit,seshadri2015fast,flashcosmos,seshadri2019dram,seshadri2016processing,seshadri2017simple,seshadri.arxiv16,li2016pinatubo,xin2020elp2im,angizi2019redram}. In \pum{}, computation takes place inside the memory (e.g., DRAM) by leveraging the analog intrinsic behavior of memory devices, resulting in reduced data movement costs.

DRAM is a prevalent main memory technology that enables \pum{} \nb{in} various systems. Prior works demonstrate that \pum{} operations in off-the-shelf DRAM chips have the potential to improve the performance and energy efficiency of commodity systems greatly~\cosDRAMPumAllCitations{}. These works enable many fundamental \pum{} operations in DRAM chips, including but not limited to \one{} bitwise arithmetic \nb{and logic} operations using three-input majority function (\maj{3})~\cite{gao2019computedram,gao2022frac}, \two{} bulk-data copy operations at DRAM row granularity~\cite{gao2019computedram} (known as RowClone~\cite{seshadri2013rowclone}), and \three{} security primitives \nb{(e.g., in-DRAM true random number generation (TRNG)~\cite{kim2019drange, olgun2021quactrng} and physical unclonable functions (PUF)~\cite{kim2018dram}}.

\subsection{Limitations of State-of-the-Art} 
We identify two key \nb{limitations} of prior work for \pum{} operations in commodity DRAM chips: \one{} success rate and \two{} \iey{low throughput and high latency}.

\subsubsection{Success Rate}
\label{subsubsec:motiv_success}

{We define \emph{the success rate} of a \maj{} operation per row group as the percentage of the bitlines that \iey{reliably} produce the correct output.
To analyze the success rate of the \maj{3} operation \agy{in} off-the-shelf DRAM chips, we 
\agy{conduct} \maj{3} experiments using the state-of-the-art mechanism\agy{: FracDRAM}~\cite{gao2022frac} on \nb{12} modern off-the-shelf DRAM modules from SK Hynix, following the methodology in \secref{subsubsec:maj_success}.

\figref{fig:four_rows} shows \agy{FracDRAM's \maj{3}}-success-rate \agy{distribution across different row groups (y-axis) for different DRAM modules (x-axis) in a box-and-whiskers plot}.\agy{\footnote{\label{fn:boxplot}\agy{{A box-and-whiskers plot emphasizes the important metrics of a dataset’s distribution. The box is lower-bounded by the first quartile and upper-bounded by the third quartile.
The \gls{iqr} is the distance between the first and third quartiles (i.e., box size).
Whiskers show the minimum and maximum values.}}}} 
\nb{The red dashed line represents the reported average success rate of \maj{3} in DDR3 modules~\cite{gao2022frac}.} \nb{We make \param{two} key observations based on \figref{fig:four_rows}.} \nb{First, \agy{FracDRAM} \nb{has a} low \nb{average} success rate \agy{
of}} \param{78.85\%} \nb{across \agy{all tested}} DRAM chips.
\nb{Second, \agy{FracDRAM's \maj{3} success rate significantly reduces (\param{19.37\%} on average) across newer generations of DRAM chips from DDR3 to DDR4.}}
\nb{Based on this observation, we expect the success rate of \maj{3} operations to reduce even more as DRAM continues to scale down in newer generations (e.g., DDR5).}

\agy{We conduct SPICE simulations to investigate} the reasons behind \maj{3}'s low success rate across different DRAM modules \agy{following the methodology in \secref{subsec:inp_repl}}. \agy{We analyze the effect of manufacturing process variation on \maj{3}'s success rate for all the possible inputs (i.e., (0,0,0) to (1,1,1)). To do so, we conduct a Monte Carlo analysis over $10^4$ iterations, where we inject \param{10}\SI{}{\percent}, \param{20}\SI{}{\percent}, \param{30}\SI{}{\percent}, and \param{40}\SI{}{\percent} variation to \yct{capacitor and transistor} parameters.}

\begin{figure}[ht]
\centering
\includegraphics[width=0.8\linewidth]{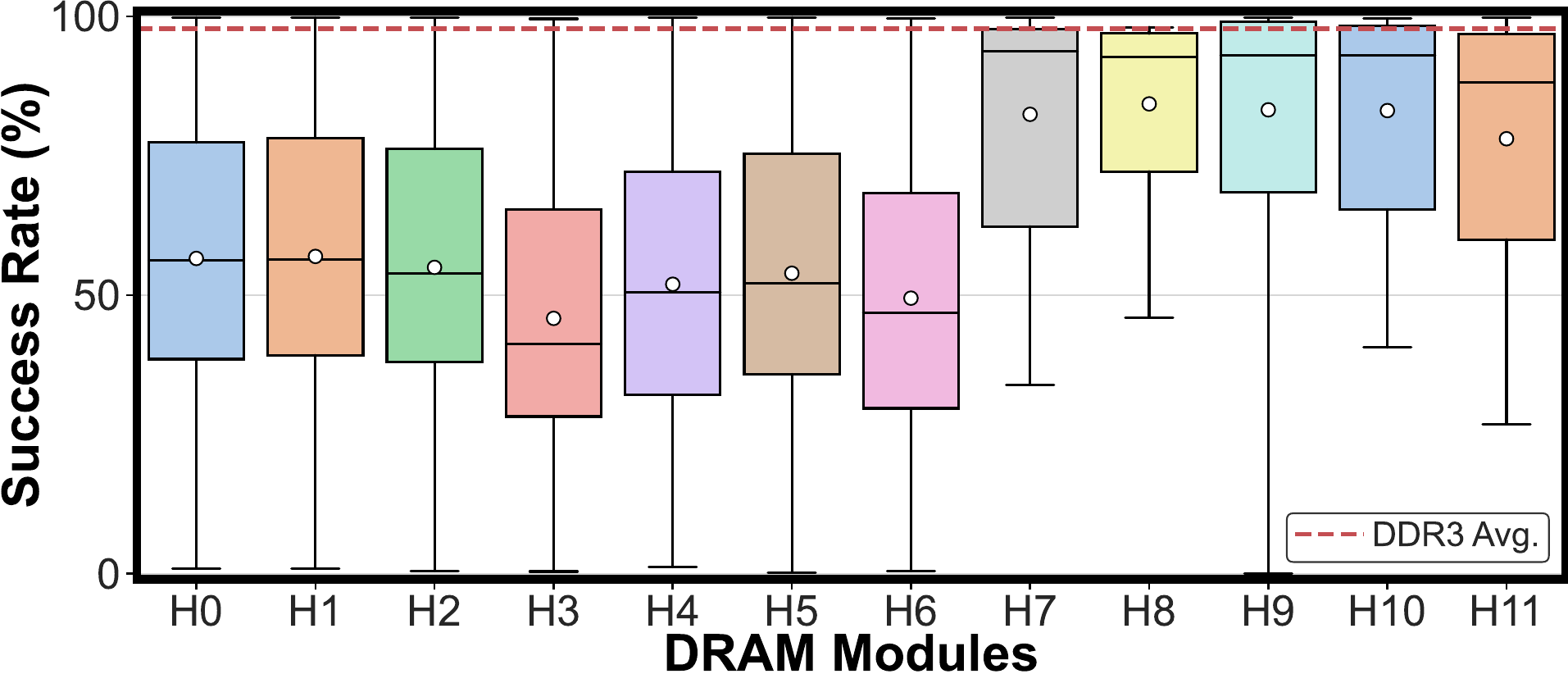}
\caption{Distribution of the \maj{3} success rate of state-of-the-art mechanism across 12 off-the-shelf DDR4 chips.}
\label{fig:four_rows}
\end{figure}

\figref{subfig:spice_sr} shows \agy{how manufacturing process variation (colors) affects \maj{3}'s success rate (y-axis) with different input patterns (x-axis) based on our} \nb{SPICE} simulation results. 
We make two key observations from \figref{subfig:spice_sr}. First, in all 1's and all 0's input patterns (i.e., (0,0,0) and (1,1,1)), \maj{3} works with a 100\% success rate as all activated cells in a bitline try to pull the bitline to the same voltage level, resulting in safe sensing operation. Second, \maj{3} operations that have at least one different value in the input data pattern (i.e., (0,0,1) to (1,1,0)) produce incorrect results with an increasing trend as the process variation percentage increases, up to \param{46.58}\%. This is because some of the activated cells attempt to pull the bitline to a level, whereas the others attempt to pull the bitline to the opposite level. Depending on the cell's characteristics, this operation produce incorrect results and thus lowers the success rate of the \maj{3}. To investigate further, we analyze the distribution of the bitline deviation when four rows are simultaneously activated to perform \maj{3} with an input pattern that has two logic-1 (e.g.,\maj{3}(1,1,0)).

\begin{figure}[ht]
\centering
\captionsetup[subfigure]{justification=centering}
\subfloat[]{{\includegraphics[width=0.5\linewidth]{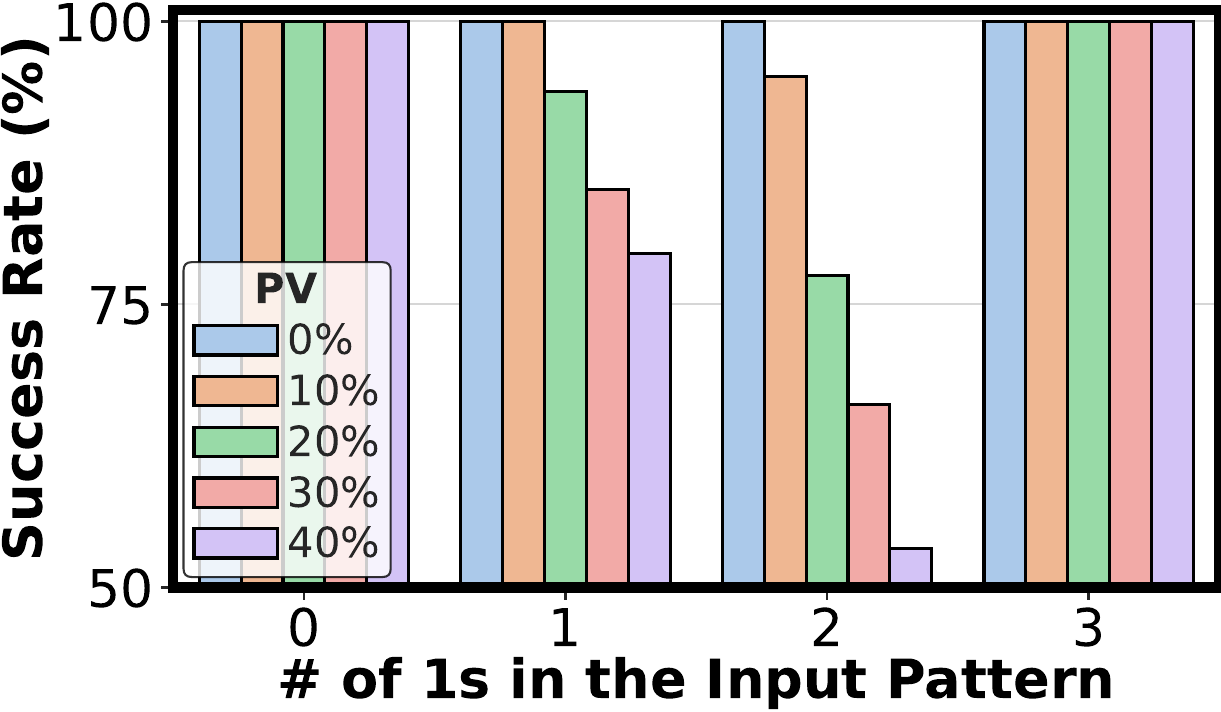}}\label{subfig:spice_sr}}
\subfloat[]{{\includegraphics[width=0.5\linewidth]{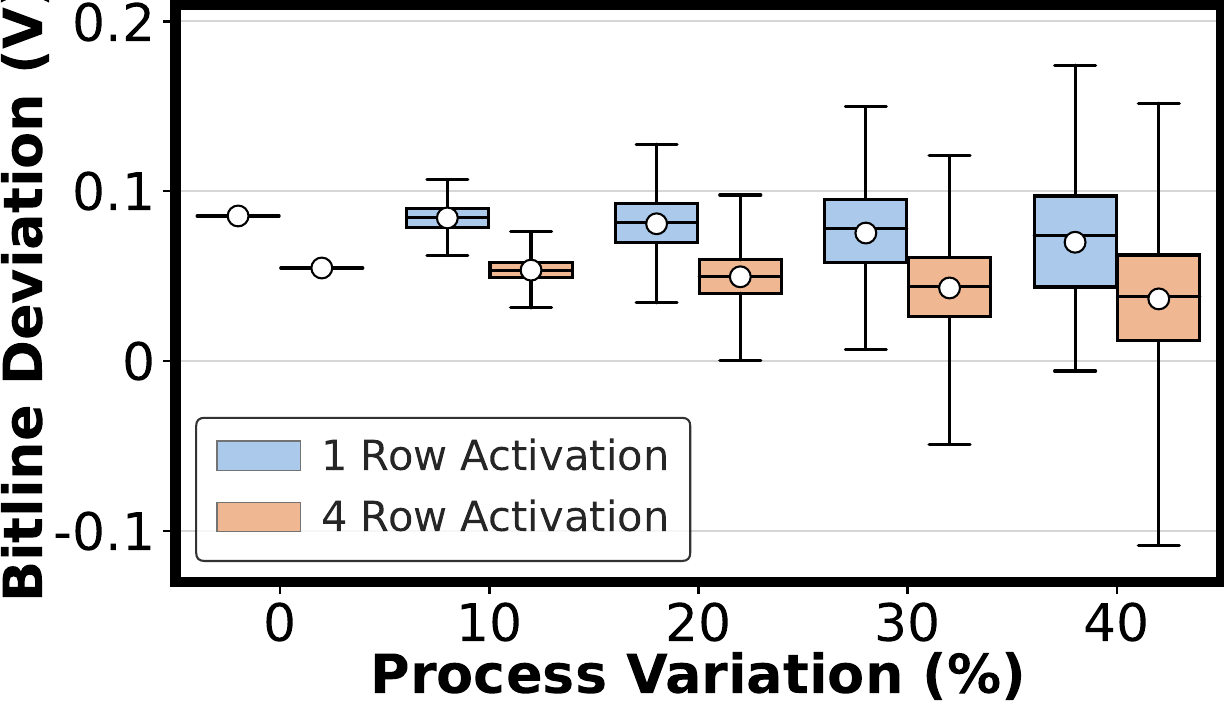}}\label{subfig:spice_dist}}
\caption{The success rate of \maj{3} in different input patterns (a) and the distribution of the bitline deviation (b) for various process variations.}
\label{fig:four_spice}
\end{figure}

\figref{subfig:spice_dist} \agy{presents a box-and-whiskers plot}\footref{fn:boxplot}
\agy{that demonstrates the effect of manufacturing process variation (x-axis) on the bitline voltage's net deviation (y-axis) when four rows are simultaneously activated to perform \maj{3}(1,1,0).}
\agy{As a comparison point,} we evaluate the deviation on the bitline when a single row \yct{that stores 1 (i.e., \vdd{})} is activated (i.e., nominal activation operation) for the corresponding process variation percentages. We make two key observations \agy{from \figref{subfig:spice_dist}}. First, activating multiple rows to perform \maj{3} with two logic-1 input patterns reduces the bitline deviation by \param{41.14}\% on average, compared to activating a single row. This is because the activated cells store conflicting data (i.e., not all 1s or all 0s), thus trying to pull bitlines to opposite voltage levels. 
Second, \agy{manufacturing} process variation significantly affects the deviation on the bitline voltage distribution, \agy{i.e., boxes get wider as the process variation increases from left to right in \figref{subfig:spice_dist}. Increased variation can cause \maj{3} operation to compute an incorrect result.} This is because process variation can cause variations in cell capacitance and affect the behavior of transistors and bitlines, as well as the latency of wordline assertion. We conclude that these variations can affect the success rate of the charge-sharing operation and, in turn, the correctness of its results.

\subsubsection{Performance}

\agy{\pum{} operations in off}-the-shelf-DRAM \agy{chips}~\cite{gao2019computedram,gao2022frac} are limited in functionality \agy{by \emph{only} \maj{3}~\cite{gao2019computedram,gao2022frac,seshadri2017ambit} and RowClone~\cite{seshadri2013rowclone,gao2019computedram} operations, }which \agy{requires them to execute complex procedures by sequentially performing these operations many times and thus hinders their} performance \agy{benefits}. 
For instance, \one{} to perform more than two-operand AND/OR operations, prior works need to perform multiple \maj{3} operations since \maj{3} can perform only two-operand AND/OR operations and \two{} to initialize \emph{N} rows, \emph{N} RowClone operations \agy{are needed as each of them} can initialize only one row at a time. \yct{These} limitations lead to reducing the potential advantages of \pum{} \agy{operations in off-the-shelf DRAM chips}.

\nb{Increasing the number of operands in \maj{} operations (e.g., \maj{5}) can significantly improve the} throughput of many applications, such as data analytics~\cite{seshadri2017ambit,jun2015bluedbm,torabzadehkashi2019catalina}, databases~\cite{wu2005fastbit, wu1998encoded, guz2014real}, and graph processing~\cite{li2016pinatubo,besta2021sisa,hajinazar2020simdram}. \nb{To demonstrate the potential benefits of enabling more than three-input \maj{}} we model 4 different \maj{} operations: \nb{\maj{3}, \maj{5}, \maj{7}, and \maj{9}.}
\nb{All operation models assume equal latency
values based on the state-of-the-art \maj{3} operation~\cite{gao2022frac} to show the potential benefit of different majority operations. 
Note that the actual latency of these operations may be higher than what is assumed in this evaluation. 

}

\figref{fig:motivation_speedup} \agy{shows the performance} speedup of operations that are based on \maj{5}, \maj{7}, \maj{9} over the \maj{3} for \agy{\one{} three bit-wise logic operations:} AND, OR, and XOR, \agy{and \two{} four bit-serial arithmetic operations }  addition (ADD), subtraction (SUB), multiplication (MUL), and division (DIV).
\agy{We implement the arithmetic operations based on full-adder designs using \maj{3} and \maj{5} operations.}
\nb{This is because the full adder design utilizes up to five-input majority operations~\cite{navi2009two, ali2019memory}.} Based on \agy{\figref{fig:motivation_speedup}}, we observe that increasing the number of operands in majority operations results in significantly \nb{higher speedups} for both arithmetic and logic microbenchmarks (e.g., \maj{9} has \param{2.73}\SI{}{\times} \nb{higher speedup over \maj{3} on average across logic microbenchmarks). Therefore, \agy{we conclude that} extending \pum{} functionality in off-the-shelf DRAM chips \agy{greatly enhances performance for} many workloads.}

\begin{figure}[ht]
\centering
\includegraphics[width=0.85\linewidth]{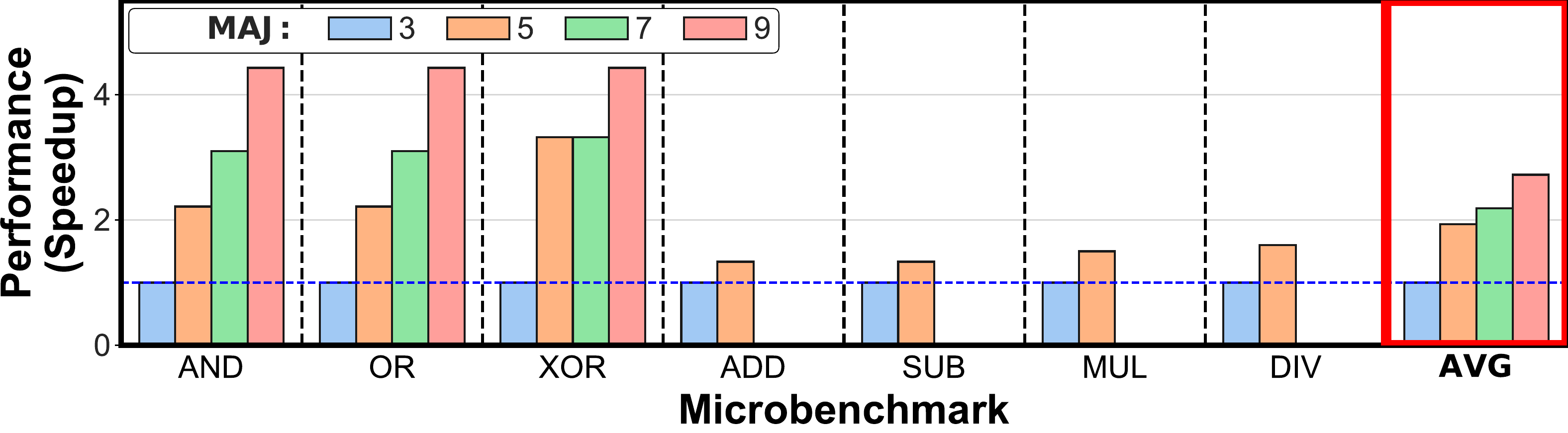}
\caption{Speedup over the \maj{3} in seven microbenchmarks.}
\label{fig:motivation_speedup}
\end{figure}

%% file: sections/04_sim-act.tex
\section{Simultaneous Many Row Activation}
\label{sec:multi-act-decoder}

We find that by carefully crafting a specific \hluo{sequence} of \gls{apa} DRAM commands with reduced timings, 2, 4, 8, 16, and 32 rows in the same subarray can be activated simultaneously. 
We characterize \nCHIPS{} modern off-the-shelf DRAM chips from two major manufacturers using an FPGA-based off-the-shelf DRAM testing infrastructure (\secref{sec:real_characterization}).
To explain the potential mechanism behind our observation, we analyze the row decoder \yct{circuitry} \hluo{of a \agy{DRAM} bank in an off-the-shelf DRAM chip}. We hypothesize that the hierarchical structure of row decoder design with multiple pre-decoding schemes allows us to simultaneously activate many rows. We present a hypothetical row decoder circuitry that explains activating many rows simultaneously (\secref{sec:reverse_engineering}).

\subsection{Real DRAM Chip Characterization}
\label{sec:real_characterization}
We demonstrate that \bralow{} works reliably on \nCHIPS{} DRAM chips that come from two major manufacturers.
Table~\ref{tab:dram_chips} provides a list of the DRAM modules along with the chip identifier (Chip ID), manufacturing date (Date), die revision (Die Rev.), chip density (Chip Dens.), and DRAM organization (ranks, banks, and pins).

\begin{figure}[ht]
\centering
\includegraphics[width=0.75\linewidth]{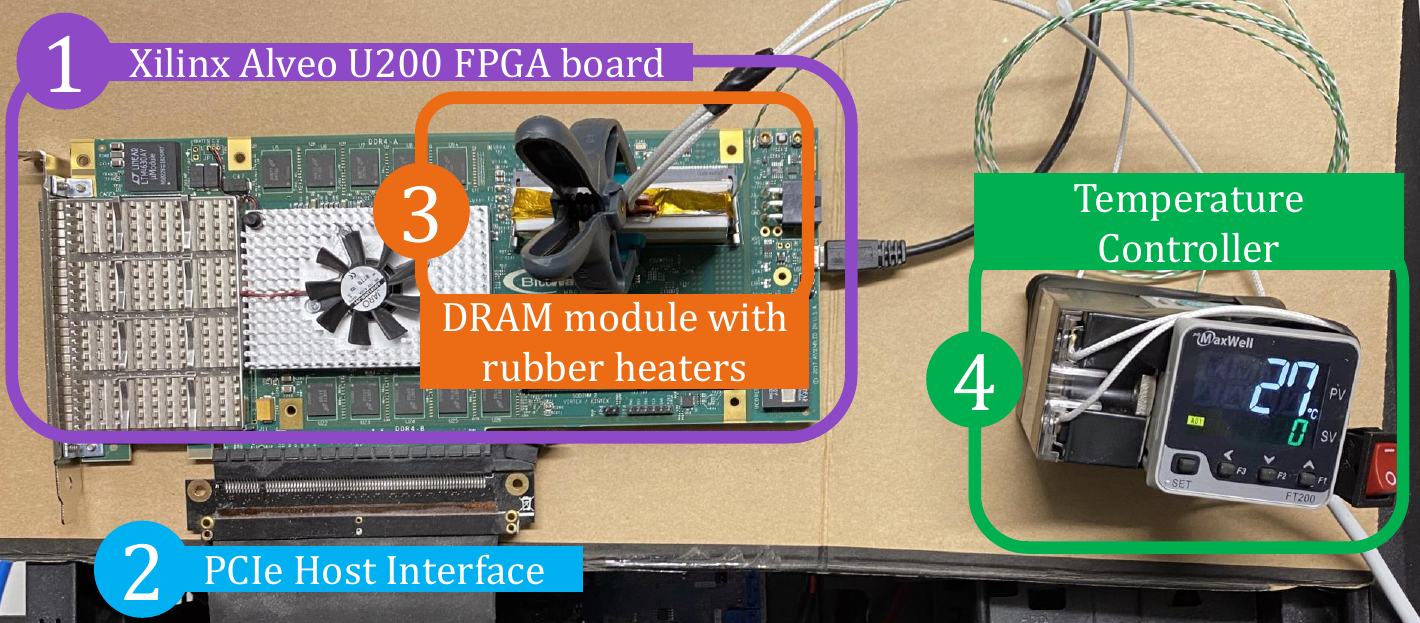}
\caption{DDR4 DRAM Bender experimental setup.}
\label{fig:infra}
\end{figure}

\input{tables/dram_chips}
\noindent
\textbf{Infrastructure.} We conduct real DRAM chip experiments on DRAM Bender~\cite{safari-drambender, olgun2022drambender}, an FPGA-based DDR4 testing infrastructure that provides precise control of the DDR commands issued to a DRAM module. \figref{fig:infra} shows 
our experimental setup that consists of four main components: \one{} the Xilinx Alveo U200 FPGA board~\cite{alveo_u200} programmed with DRAM Bender \two{} a host machine that generates the sequence of DRAM commands that we use in our tests, \three{} rubber heaters that clamp the DRAM module on both sides to avoid fluctuations in ambient temperature, and \four{} a MaxWell FT200~\cite{maxwellFT200} temperature controller that controls the heaters and keeps the DRAM chips at the target temperature.

\noindent
\textbf{Verification Experiment.} We 1) initialize N rows that would be activated simultaneously, which we referred to as the N row groups (\nrg{}), with a predetermined data pattern, 2) perform \yct{\apaLong{} command sequence}
(\apa{}) \yct{with reduced timings} on the \nrg{} to simultaneously activate multiple rows, 3) issue a \wri{} command while all rows in \nrg{} are active, and 4) precharge the bank and individually read each row in \nrg{} while adhering to the manufacturer-recommended DRAM timing parameters. If the rows are activated with \apa{} command sequence, \wri{} \yct{command} overwrites the predetermined data pattern with the new one. We observe that all rows in \nrg{} are updated with the newly written data pattern. We observe that up to 32 rows can be activated simultaneously in one major DRAM manufacturer, while up to 16 can be activated in another major DRAM manufacturer. 

\noindent
\textbf{Finding All \nrg{} in a Subarray.} 
We successfully reverse engineer the number of subarrays and subarray size (listed in Table~\ref{tab:dram_chips}) using RowClone~\cite{gao2019computedram}, which is used in many prior works to determine subarray boundaries~\cite{olgun2021pidramgithub, olgun2022pidram}. To investigate which rows are simultaneously activated in a subarray,  we perform \apaEx{} command sequence with reduced timing parameters, where the \rfirst{} is the firstly activated row and the \rsecond{} is the secondly activated row. We test every possible \rfirst{} and \rsecond{} combinations of this sequence and record the row addresses that are simultaneously activated in a subarray. We present in Table~\ref{tab:dram_chips} the percentage of the number of rows that can be activated simultaneously out of all two-row address pairs in a subarray (\nrg{}\%) across different DRAM chips and manufacturers.

\subsection{Hypothetical Row Decoder Design}
\label{sec:reverse_engineering}
The row decoder circuitry in a DRAM bank \hluo{decodes the n-bit row address (RA)} and asserts a wordline out of $2^n$ wordlines. Modern DRAM chips have multiple tiers of decoding stages (pre-decode and decode stages) to reduce latency, area, and power consumption~\cite{bai2022low,weste2015cmos,turi2008high}. We analyze the row decoder circuitry of an off-the-shelf DRAM \hluo{chip}, H8 module which has \param{$2^{16}$} rows in a bank. We observe that in H8, the subarrays consist of \param{$2^9$} rows, and the total number of subarrays in a bank is \param{$2^7$}. We present a hypothesis regarding the row decoder circuitry \yct{that allows simultaneous activation of many rows} and the sequence of operations that occur in the row decoder when \act{} and \pre{} commands are issued.

\noindent
\textbf{Row Address Indexing.}  Based on the characterization results, we hypothesize that the lower-order 9 bits of the RA are used to index the row within a subarray, while the higher-order 7 bits are used to index the corresponding subarray. 

\begin{figure}[ht]
\centering
\includegraphics[width=\linewidth]{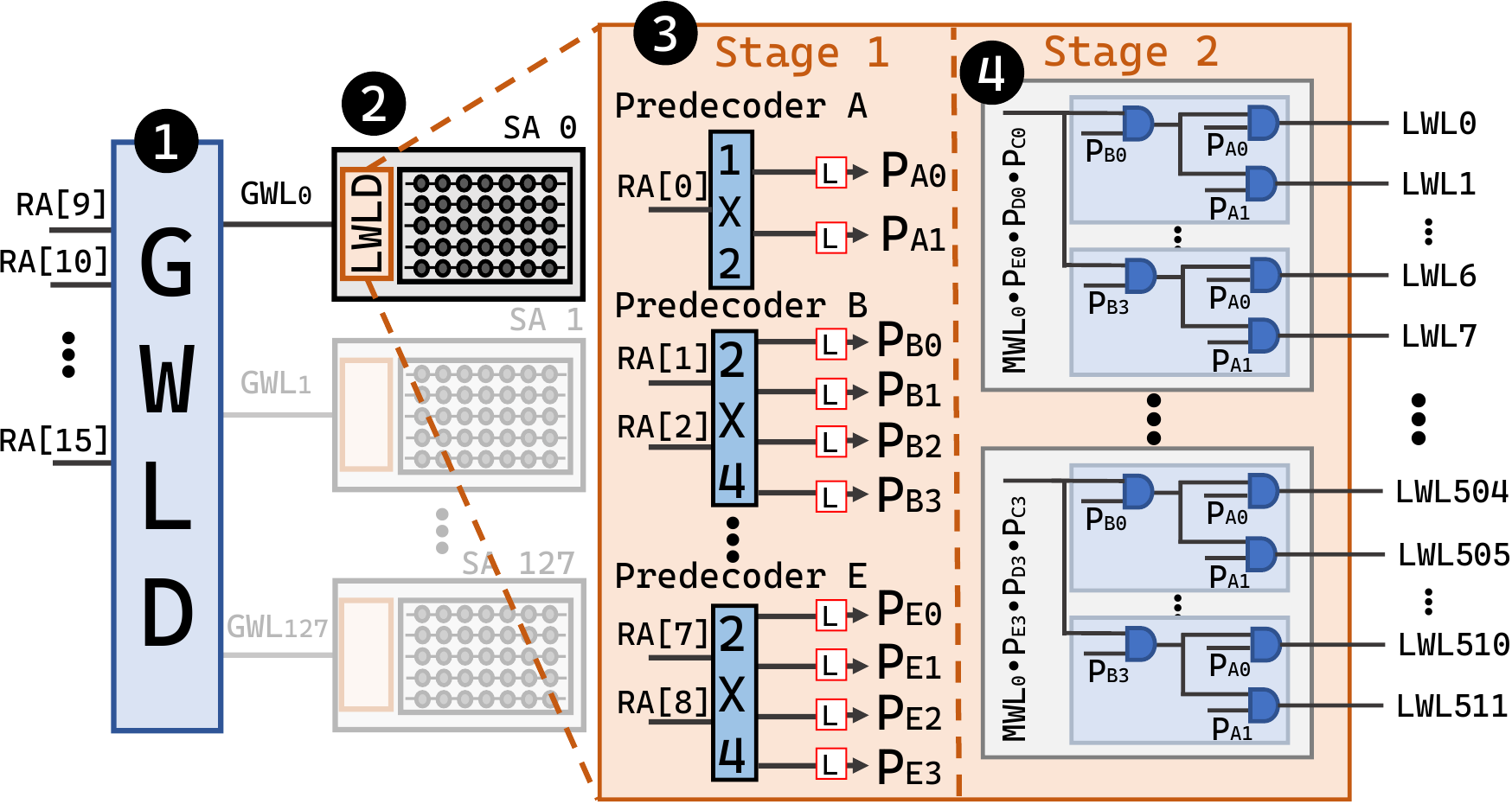}
\caption{Hypothetical Row Decoder Design.}
\label{fig:row-decoder}
\end{figure}

\noindent
\textbf{Row Decoder Design.}~\figref{fig:row-decoder} illustrates the potential row decoder circuitry of a DRAM bank in an off-the-shelf DRAM module that consists of two decoding stages: 1) Global Wordline Decoder (GWLD) (\dingOne{}) and 2) Local Wordline Decoder (LWLD) (\dingTwo{}). When an \act{} command is issued, three operations occur in order. First, GWLD decodes the higher-order 7 bits of the RA (RA[9:15]) and drives the corresponding Global Wordline ($GWL$) that is connected to the LWLD of \yct{the corresponding} DRAM subarray. Second, \yct{Stage 1 of LWLD pre}decodes the lower-order 9 bits of the RA \yct{(RA[0:8])} in five tiers of predecoders \yct{(Predecoder A/B/C/D/E, \dingThree{})} and latches the predecoded address bits \yct{($P_{A0}, P_{A1}, ..., P_{E3}$)}, a total of 18 bits. Third, \yct{Stage 2 of LWLD decodes the} predecoded $P$ signals to assert the corresponding Local Wordline ($LWL$)~\yct{in the} Stage 2, which consists of 64 sub-decoder trees (\dingFour{}). When a \pre{} command is issued, the latched predecoded \yct{$P$ signals} are reset \yct{to de-assert} the corresponding $LWL$.

\noindent
\textbf{Activating Multiple Rows: A Walk-Through.}
Reducing the latency between \pre{} and the \yct{second} \act{} commands (i.e., $t_{RP}$) prevents the reset operation and allows the predecoders to latch the next RA without deasserting the RA \yct{targeted by} the first \act{} command. Hence, after the second \act{} command, depending on the target addresses of \apa{} sequence, \yct{one or two} latches of \yct{each} pre-decoder in LWLD \yct{can be} set.
\yct{By changing the row addresses targeted by two \act{} commands, we can control the number and addresses of the simultaneously activated rows in a subarray.} 

\begin{figure}[ht]
\centering
\includegraphics[width=\linewidth]{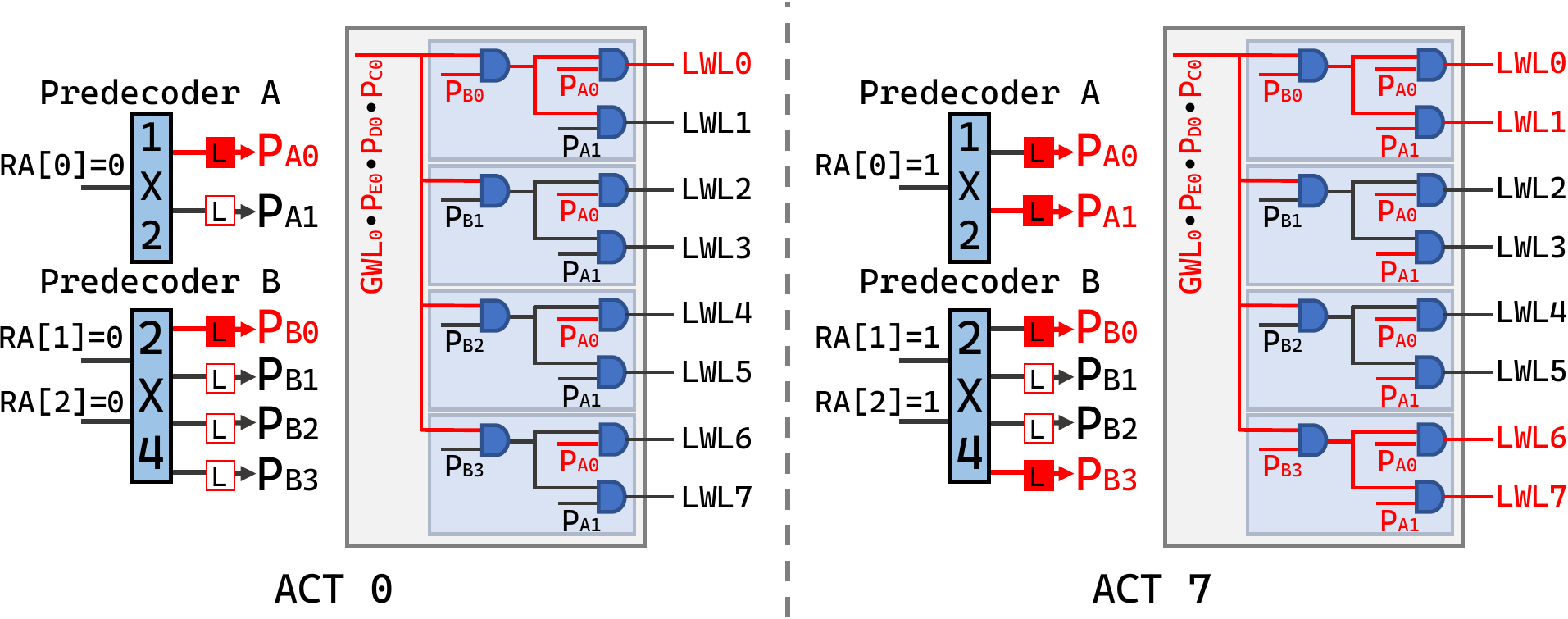}
\caption{Example of activating multiple rows in hypothetical row decoder design. The red colors represent asserted signals.}
\label{fig:mra_example}
\end{figure}

\yct{\figref{fig:mra_example} demonstrates an example of how the hypothetical row decoder design enables simultaneously activating four rows in the same bank
when an \apa{} command sequence targeting Row 0 ($...0000_2$) and Row 7 ($...0111_2$) (i.e., \apaex{}) is issued.}
\yct{When the first \act{} 0 is received, the predecoders assert $P_{A0}$ and $P_{B0}$ signals. \agy{The asserted $P_{A0}$ and $P_{B0}$ signals drive $LWL_{0}$.}}
When the \act{} 7 is received, \agy{the predecoders \agy{assert} $P_{A1}$ and $P_{B3}$ signals. Due to issuing \act{} 7 command with} reduced timings, \agy{the signals $P_{A0}$ and $P_{B0}$ are not yet de-asserted, and thus all of}
$P_{A0}$, $P_{A1}$, $P_{B0}$, and $P_{B3}$ signals are set simultaneously, \agy{and thus the} decoder tree asserts \agy{all of} $LWL_{0}$, $LWL_{1}$, $LWL_{6}$, and $LWL_{7}$ \agy{wordlines}, \agy{thereby} simultaneous\agy{ly} activati\agy{ng all of rows 0, 1, 6, and 7.} 

\begin{figure}[ht]
\centering
\includegraphics[width=0.9\linewidth]{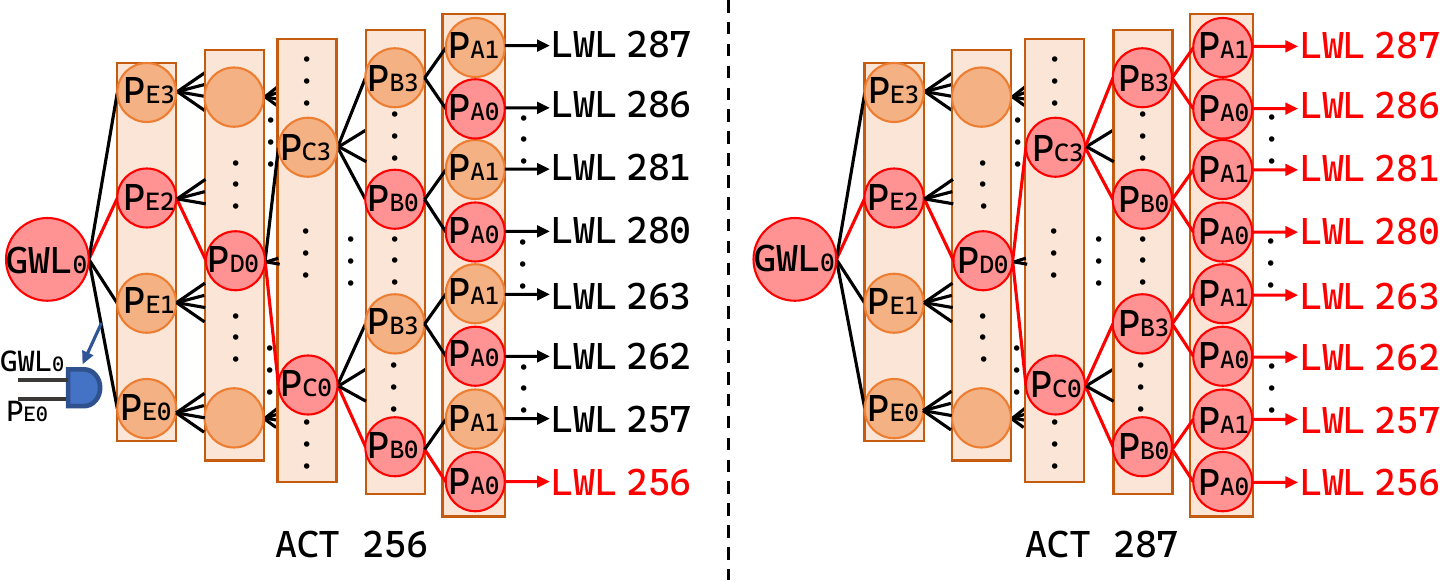}
\caption{A high-level abstraction of row decoder tree. The red colors represent asserted signals.}
\label{fig:graph_ex}
\end{figure}

\figref{fig:graph_ex} depicts a higher-abstraction level for the hierarchical row decoder tree in the first subarray when an \apa{} command sequence targets Row 256 and Row 287. Each node represents a signal that is used in the decoding process. The first (the root) node is the output of GWLD ($GWL_{0}$), other nodes are the predecoded address signals \yct{($P_{A0}, P_{A1}, ..., P_{E3}$)}. Each edge between nodes represents the AND gate of the nodes. Each box represents the predecoders E/D/C/B/A (starting from root to leaf), which is the level of the row decoder tree. When we issue \act{} 256, it is decoded through the circuitry and asserts the corresponding predecoded address signals \yct{($P_{E2}$, $P_{D0}$, $P_{C0}$, $P_{B0}$, $P_{A0}$)}, highlighted as red on the left side of the figure. When the \act{} 287 is issued with the reduced timings, $P_{C3}$, $P_{B3}$, and $P_{A1}$ are \yct{also} latched, resulting in activating eight rows in a subarray.

We can formulate our observation as follows: to activate $2^N$ rows, N different predecoders have to latch two signals. For instance, to activate four rows, we issue \apa{} commands by targeting the rows that only latch the two outputs of \yct{two different predecoders} 
as illustrated in \figref{fig:mra_example}. Hence, to activate thirty-two rows, \apa{} command sequence needs to target such rows that make all predecoders latch two outputs (e.g., \act{} 127 $\rightarrow$ \pre{} $\rightarrow$ \act{} 128). We hypothesize that the upper bound for the number of rows that are simultaneously activated depends on the number of predecoders. The examined module has five predecoders, and thus we activate up to $2^5$ rows.

%% file: tables/dram_chips.tex
\begin{scriptsize}
\begin{table*}[ht]
\centering

\resizebox{\textwidth}{!}{%
\begin{tabular}{@{}lllccrccccccccc@{}}
\multirow{2}{*}{{{\bf Manufacturer}}} & \multirow{2}{*}{\textbf{Module}} & \multirow{2}{*}{{{\bf  Chip ID}}} & {\bf Date} & {{{\bf Die}}} & {{{\bf Chip}}} & \multicolumn{3}{c}{{{\bf Organization}}} & {\bf SA} & \multicolumn{5}{c}{{\bf \nrg{}\%}}  \\
\cmidrule(lr){7-9} \cmidrule(lr){11-15}
 &  &  & (yy-ww) & \textbf{Rev.} & \textbf{Dens.} & \textit{Ranks} & \textit{Banks} & \textit{Pins} & \textbf{Size} &  \textit{2} & \textit{4} & \textit{8} & \textit{16} & \textit{32} \\ \midrule \midrule
SK Hynix & H0-6  & H5AN4G8NMFR & Unknown   & M & 4Gb  & 1 & 16 & x8  &  512-640 & \SI{2.07}{\percent} & \SI{10.65}{\percent} & \SI{25.37}{\percent} & \SI{26.81}{\percent} & \SI{9.91}{\percent} \\
(Mfr. H) & H7-11 & H5AN4G8NAFR & Unknown   & A   & 4Gb  & 1 & 16 & x8  &  512 & \SI{2.49}{\percent} & \SI{12.63}{\percent} & \SI{30.77}{\percent} & \SI{35.33}{\percent} & \SI{1.83}{\percent} \\ \midrule
{Micron} & M0-3  & OUE75 D9ZFW & 20-46 & E   & 16Gb & 1 & 16 & x16 & 1024 & \SI{1.91}{\percent} & \SI{12.92}{\percent} & \SI{32.87}{\percent} & \SI{20.83}{\percent} & \SI{0 }{\percent} \\
(Mfr. M) & M4-5  & 1LB75 D9XPG & 21-26   & B   & 16Gb & 1 & 16 & x16 & 1024 & \SI{1.47}{\percent} & \SI{8,11}{\percent} & \SI{15.27}{\percent} & \SI{11.06}{\percent} & \SI{0 }{\percent} \\ \bottomrule \bottomrule
\end{tabular}%
}
\caption{Summary of DDR4 DRAM chips tested.}
\label{tab:dram_chips}

\end{table*}
\end{scriptsize}

%% file: sections/05_mechanism.tex
\section{\X{}}
\label{sec:multi-act}

\X{} leverages \bralow{} and demonstrates a proof-of-concept to improve success rate and the performance of \pum{} operations in off-the-shelf DRAM chips by \one{} replicating the inputs and \two{} introducing new \pum{} primitives.

\subsection{Input Replication}
\label{subsec:inp_repl}

\agy{Although modern} off-the-shelf DRAM chips do not officially support \maj{3}\agy{, it is possible to} perform \maj{3} \agy{operation} in off-the-shelf DRAM chips \agy{on four simultaneously activated rows} 
by violating \agy{two} timing parameters\agy{: \gls{tras} and \gls{trp}~\cite{gao2022frac}.}
This mechanism can reduce the deviation on the bitline voltage, depending on the data that are stored in activated cells (\figref{subfig:spice_dist}), making it highly susceptible to electrical noise and process variation. Hence, state-of-the-art mechanism-based~\cite{gao2022frac} \maj{3} operations suffer from a low success rate. \nb{To improve the low success rate of \maj{3} operations,} \X{} increases the deviation on the bitline voltage towards safe sensing margins. \agy{\X{} achieves this by} storing multiple copies of each input on all simultaneously activated rows, i.e., replicating the input operands.

\begin{figure}[ht]
\centering
\includegraphics[width=0.68\linewidth]{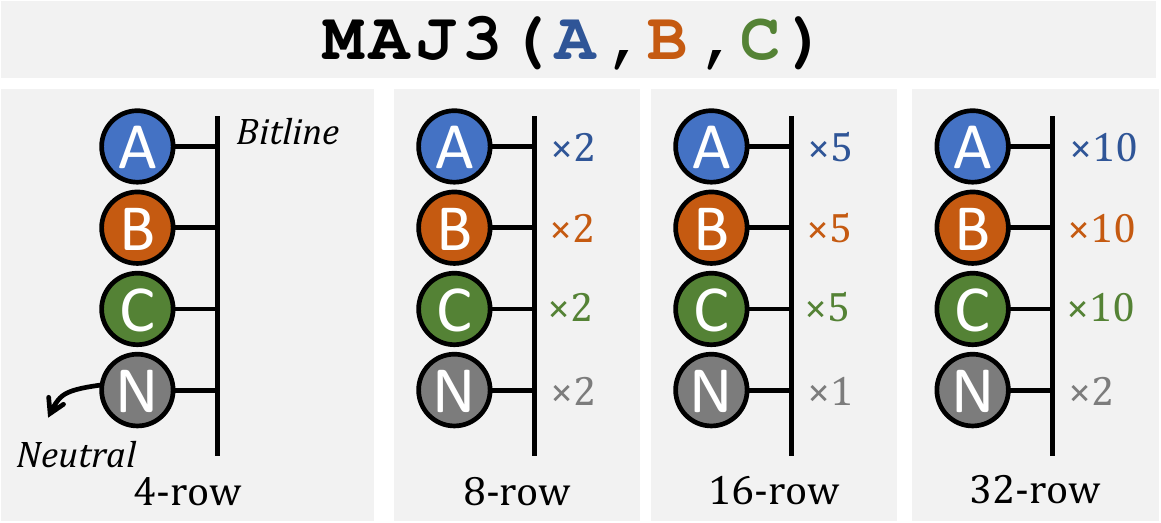}
\caption{\nb{\maj{3}(A, B, C) utilizing 4-, 8-, 16-, 32-row activation with input replication}.}
\label{fig:repl_ex}
\end{figure}

\nb{Input replication exploits the majority Boolean algebra rule, where replicating input operands maintains the functionality (e.g., \maj{6}(A,B,C,A,B,C) = \maj{3}(A,B,C)). \figref{fig:repl_ex} illustrates \maj{3}(A, B, C) utilizing 4-, 8-, 16-, 32-row activation with input replication. The state-of-the-art 4-row activation-based \maj{3}, FracDRAM~\cite{gao2022frac} places one row in the neutral state (N) while activating four rows simultaneously. For \yct{N}-row activation-based \maj{3} with \yct{N}$ > 4$, all inputs are replicated to the maximum extent possible. The remaining rows are then set to the neutral state.}

\begin{figure}[ht]
\centering
\captionsetup[subfigure]{justification=centering}
\subfloat[]{{\includegraphics[width=0.5\linewidth]{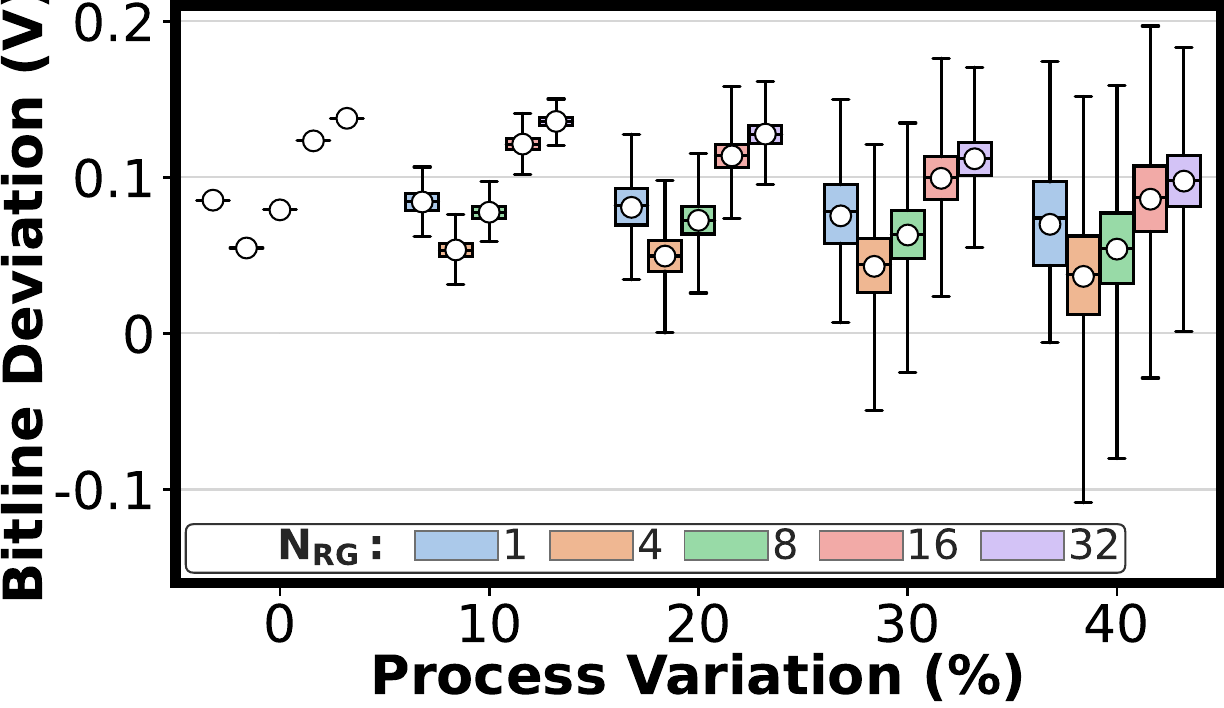}}\label{subfig:repl_spice2}}
\subfloat[] {{\includegraphics[width=0.5\linewidth]{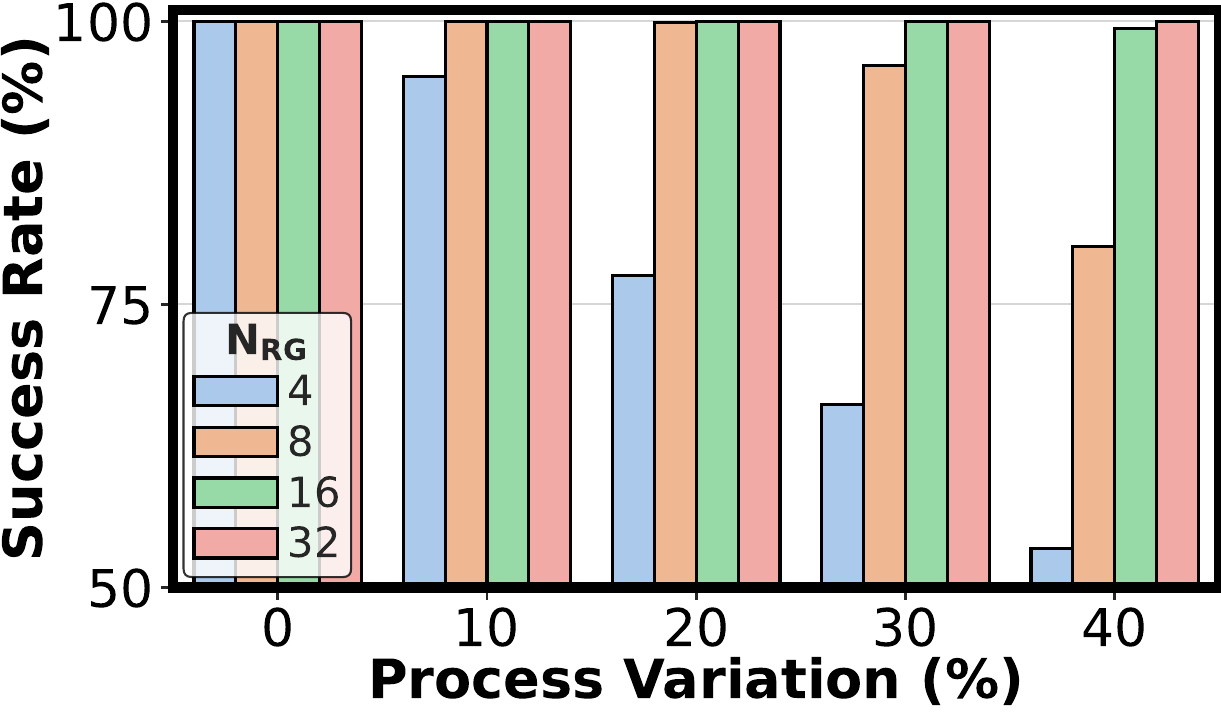}}\label{fig:spice_all}}
\caption{The effect of input replication bitline deviation (a) and the success rate of \maj{3} (b) for various \nrg{} across different process variations using SPICE simulations.}
\label{fig:repl_spice}
\end{figure}

We hypothesize that by leveraging input replication, \X{} increase\nb{s} the deviation on the bitline voltage towards the safe thresholds and, thus, reduces the effect of process variation. To study our hypothesis, we conduct SPICE simulations and analyze the effect of input replication on the success rate of the sensing operation for \maj{3}(1,1,0) operations. We use the reference 55 nm DRAM model from Rambus~\cite{rambus_model} and scale it based on the ITRS roadmap~\cite{itrs_model,vogelsang2010understanding} to model the 22 nm technology node following the PTM transistor models~\cite{ptm2012transistors}.
\nb{\figref{fig:repl_spice} shows the effect of process variation on the sensing operation when N rows are activated (where $N \in\{1,4,8,16,32\}$) simultaneously.}
\figref{subfig:repl_spice2} depicts \yct{the deviation on the bitline voltage distribution (y-axis) for different process variation percentages (x-axis).}
\nb{Each \nrg{}$=1$ box represents the bitline voltage deviation distribution for a single row activation. Boxes for other \nrg{} values show the bitline voltage deviation distribution for 4-, 8-, 16-, and 32-row activation scenarios.}
\nb{\figref{fig:repl_spice}b shows the success rate corresponding to the \maj{3} operations based on N-row activation, where $N \in\{4,8,16,32\}$.}

We make three key observations based on \figref{fig:repl_spice}. First, increasing the number of rows that are simultaneously activated increases the deviation on the bitline voltage in every process variation percentage. On average, using thirty-two rows to perform \maj{3} (i.e., ten copies for each operand and two neutral rows) have \param{159.05}\% higher deviation voltage than the FracDRAM. Second, activating more than eight rows always results in a higher deviation voltage than single-row activation on average for every process variation percentage. 
\nb{Third, input replication results in a higher success rate under \yct{all} process variation \yct{percentages}. Increasing process variation results in a lower success rate for} \maj{3} \nb{operations with less or no input replication, such as} \maj{3} \nb{with 4-row activation. The success rate of} \maj{3} \nb{based on 4-row activation reduces by \param{46.58}\% when process variation increases from 0\% to 40\%. In contrast, the success rate of }\maj{3} \nb{with 32-row activation reduces only by \param{0.01}\%.} 
We conclude that input replication increases the deviation on the bitline voltage towards the safer sensing margins and reduces the effect of process variation on \maj{3} operation's success rate.

\subsection{\nb{New \pum{} Primitives}}
\nb{\X{} introduces new \pum{} primitives enabled by \bralow{}: \mrc{}, many-input charge-sharing operations, and Bulk-Write. These new \pum{} primitives improve the performance of \pum{} techniques in off-the-shelf DRAM chips.} For all the examples that describe the compute primitives, assume activating an arbitrary row \rfirst{}, precharging and activating another arbitrary row \rsecond{} (\apaEx{}) activates eight rows simultaneously.

\subsubsection{\mrc{}}

\begin{figure}[ht]
\centering
\includegraphics[width=0.85\linewidth]{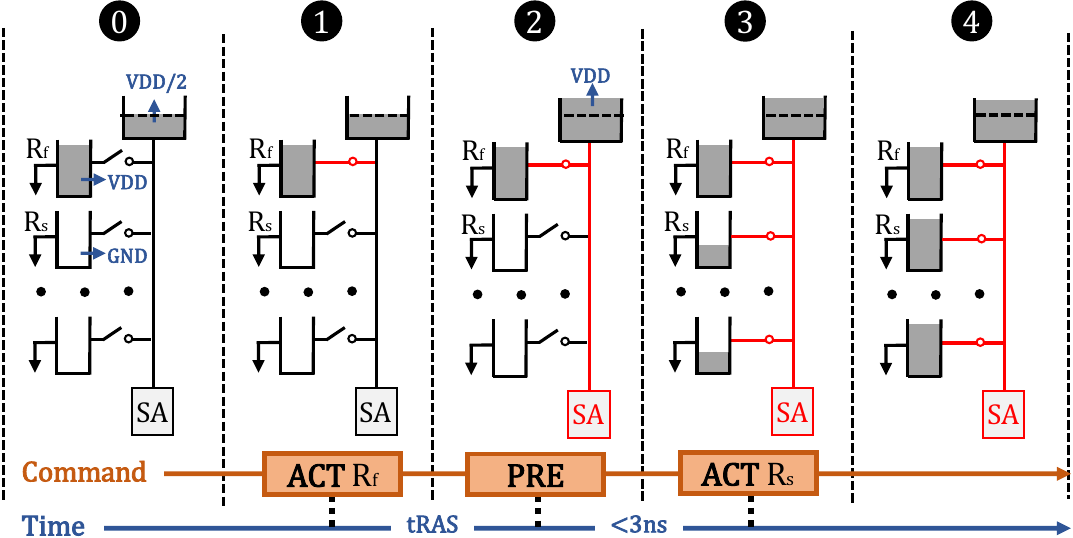}
\caption{\mrc{}.}
\label{fig:mrc}
\end{figure}

\nb{\mrc{} copies the content of a row to multiple different rows at once.}  
\iey{\figref{fig:mrc} demonstrates how the content of \rfirst{} is copied to eight rows by issuing the \apaEx{} command sequence that activates eight rows simultaneously}. Initially, the cells in \rfirst{} are charged to \vdd{}, while the other rows are at \gnd{} (\dingZero{}).

\mrc{} operation consists of \nb{four} steps. 
\nb{First, \X{} issues \act{} \rfirst{} to assert the wordline and to connect \rfirst{} to the bitline (\dingOne{}).}
\nb{Second, \X{} issues \pre{} after \gls{tras}. By obeying the \gls{tras} parameter, \X{} ensures the sense amplifier senses the \rfirst{} correctly and drives bitlines to the \rfirst{}'s charge, \texttt{VDD} (\dingTwo{}).}
\nb{Third, \X{} issues \act{} by violating \gls{trp}.} The last \act{} command interrupts the \pre{} command. By doing so, \X{} 1) prevents the bitline from being precharged to \vddh{}, 2) keeps \rfirst{} and the sense amplifier enabled, and 3) simultaneously enables the remaining seven rows (\dingThree{}). Finally, since this mechanism keeps the sense amplifier enabled that already latched the content of \rfirst{}, all activated rows are fully charged to \rfirst{} data, \vdd{} (\dingFour{}).

Leveraging \mrc{} primitive, \X{} can copy one row's data to $2^n$ rows by simultaneously activating $2^n$ rows, where $n \in\interval{1}{5}$ in off-the-shelf DRAM chips.

\subsubsection{\nb{Many-Input Charge-Sharing}}
\label{sec:charge_sharing}

\begin{figure}[ht]
\centering
\includegraphics[width=0.85\linewidth]{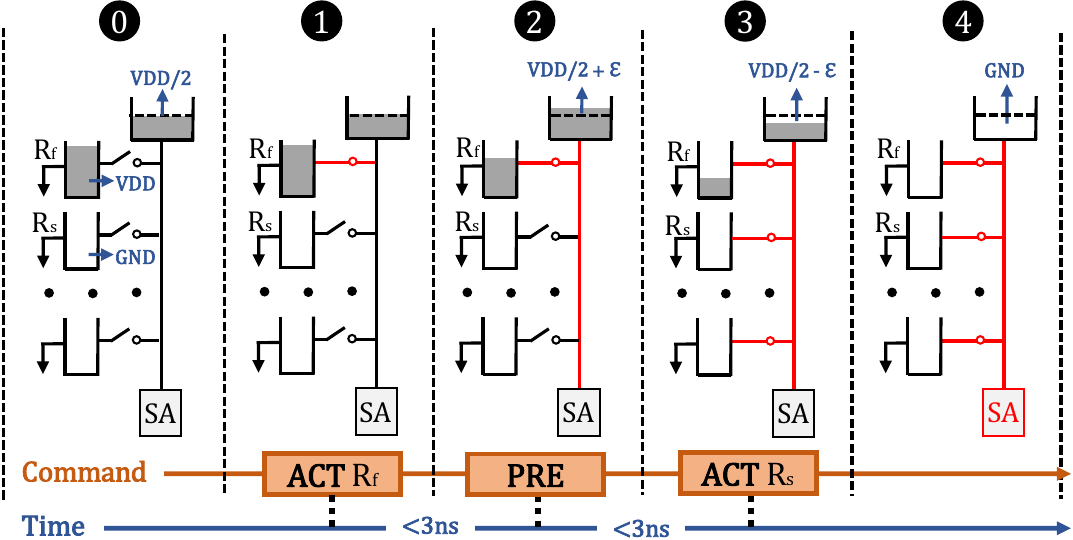}
\caption{Many-input Charge Sharing.}
\label{fig:charge_sharing}
\end{figure}

\X{} utilizes a many-input charge-sharing mechanism to extend the off-the-shelf-DRAM-based \pum{} functionality. \figref{fig:charge_sharing} depicts the many-input charge-sharing mechanism \iey{that performs eight-input majority operation by} issuing the \apaEx{} command sequence to activate eight rows simultaneously. Initially, \rfirst{} is charged to \texttt{VDD}, while the remaining rows activated by the command sequence are at \texttt{GND} (\dingZero{}). 

The many-input charge-sharing mechanism consists of four steps. 
First, \X{} issues an \act{} command to assert the wordline of \rfirst{} and thus connects \rfirst{} to the bitline (\dingOne{}). 
Second, \X{} issues \pre{} command immediately after the first \act{} in $<$ 3ns. \nb{This way, the \rfirst{} does not have sufficient time to share its charge fully} with bitline (\dingTwo{}). Third, \X{} sends the last \act{} command by greatly violating \gls{trp} ($<$ 3ns), which prevents de-asserting the \rfirst{} and activates the remaining seven rows, making all eight rows share their charge with the bitline and resulting in an eight-input majority operation. Since the majority of the activated cells have \gnd{} in this example, this leads to a negative deviation on the bitline (\dingThree{}). Finally, the sense amplifier amplifies the negative deviation and drives bitline to \texttt{GND}, leading to full discharge to all eight rows (\dingFour{}).

Leveraging many-input charge-sharing mechanism, \X{} extends the functionality of in-DRAM operations by enabling ($2n-1$)-input majority operations where $n \in\interval{2}{16}$ in off-the-shelf DRAM chips by simultaneously activating up to 32 rows. \X{} utilizes the prior work's compute primitive~\cite{gao2022frac} to perform majority operations when an even number of rows are simultaneously activated by neutralizing rows in the charge-sharing process. For instance, in the \figref{fig:charge_sharing}, putting 1) three rows into a neutral state enables \maj{5}, and 2) one row into a neutral state enables \maj{7} operation.

\subsubsection{Bulk-Write Mechanism}

\X{} introduces a compute primitives that writes data to multiple rows at once, which we call the Bulk-Write. \X{} performs the Bulk-Write operation in two steps. First, \X{} issues \apaEx{} to perform charge-sharing among eight rows using the mechanism from \secref{sec:charge_sharing}. Second, \X{} issues a \wri{} command to write data to all activated rows in a single operation. Since the activated rows are connected to bitline, \wri{} command drives the bitline to the input data, making all activated rows overwrite their data and storing the input data from \wri{} command. Leveraging this mechanism, \X{} greatly extends the multiple write operations into one Bulk-Write operation. The Bulk-Write operation can be extended to write data to up to $2^n$ rows simultaneously, where $n\in\interval{1}{5}$.

%% file: sections/06_use-cases.tex
\section{Use Cases}
\label{sec:use-cases}
This section presents our characterization and evaluation of \nCHIPS{} real DRAM chips using the infrastructure described in \secref{sec:real_characterization}. We demonstrate the effectiveness of \X{} on two\footnote{We believe that \X{} can be leveraged from other \pum{} operations. We discuss other potential use cases in \secref{subsec:extending}.} fundamental off-the-shelf-DRAM-based \pum{} use cases: 1) majority-based computation and 2) cold-boot-attack defense. 

 We demonstrate that \X{} 1) significantly increases the success rate of \maj{}, and 2) achieves significant performance gain over the state-of-the-art mechanisms.

\input{sections/06-1_maj}

\input{sections/06-3_bulk-init}

%% file: sections/06-1_maj.tex
\subsection{Majority Operation}
\label{subsec:maj}
We experimentally characterize many-input majority operations (denoted as \maj{M}, where $M \in\{3,5,7,9\}$) across different data patterns and the N rows activated simultaneously (denoted as \nrg{}, where $N \in\{4,8,16,32\}$), in real DDR4 chips through experimental evaluations. To our knowledge, our evaluations provide the first comprehensive effort to 1) characterize the success rate of the \maj{3} operation in real DDR4 chips and 2) demonstrate new operations, such as \maj{5} and \maj{7} with high reliability.

We evaluate \X{} using majority-based arithmetic and logic microbenchmarks. Our results show that introducing new operations leads to significant performance gains in the evaluated microbenchmarks.

\subsubsection{Success Rate of Majority Operations}
\label{subsubsec:maj_success}
We perform majority operations in off-the-shelf DRAM chips in four steps: we 1) initialize \nrg{} to perform \maj{M} with a data pattern, 2) perform Frac operation\footnote{For the Mfr. M, Frac operation is not supported. However, we observe that the sense amplifiers of these modules are always biased to one or zero (\nb{i.e.,} not random) depending on the cell polarity (i.e., true or anti). Initializing the neutral rows with all zeros/ones enables majority operation.} into one or multiple rows (depending on the \nb{values of} N and M) to make them neutral during charge-sharing, 3) execute a charge-sharing operation (described in \secref{sec:charge_sharing}) on the \nrg{}, and 4) read back the values in the row buffer. 

\noindent
\textbf{Success Rate.} We define a metric to evaluate majority operations, which we call \emph{the success rate}. Success rate refers to the percentage of bitlines (a total of \param{65536}) that produce correct output in all trials per \nrg{}. If a bitline produces an incorrect result at least once, we refer to this bitline as an \emph{unstable bitline} that cannot be used to perform majority operations. For example, if an \nrg{} has a 25\% success rate, it means a quarter of the bitlines (i.e., \param{16384} of the bitlines) are stable (i.e., produce correct results all the time) and can be used to perform majority operations. 

\noindent
\textbf{Data Pattern Dependence.} We analyze how the data patterns used in initializing \nrg{} affect the result of \maj{M} operations. We initialize rows in the \nrg{} with two different data patterns: 1) all ones/zeros pattern: either all ones or all zeros, and 2) random pattern: random data. We conduct our experiments on randomly selected 100 different \nrg{} in a subarray for three randomly selected subarrays in each bank, which \nb{results in} a total of 4.8K \nrg{} for each tested module.  We repeatedly perform the \maj{M} operation $10^4$ times for random data patterns and $2^M$ times for all ones/zeros patterns (i.e., all truth table inputs for a given M).

\begin{figure}[ht]
\centering
\includegraphics[width=0.85\linewidth]{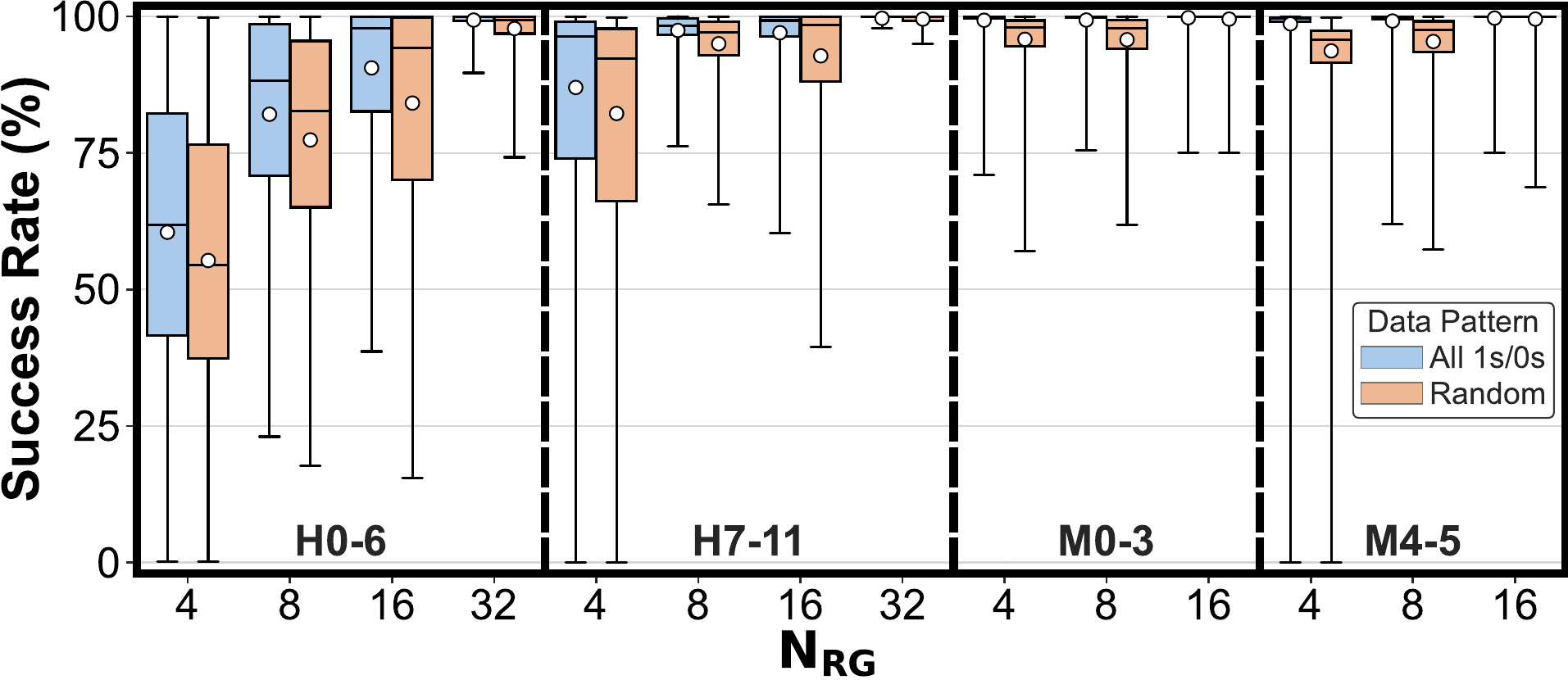}
\caption{\maj{3} Success Rate of \nrg{} for every N value across different DDR4 Modules.}
\label{fig:maj3_success}
\end{figure}

\figref{fig:maj3_success} shows a box-and-whiskers plot\footref{fn:boxplot} of the \maj{3} success rate of \nrg{} for every N value (x-axis) across different module groups. 
The state-of-the-art mechanism for \maj{3} is based on \nrg{}$=4$, FracDRAM~\cite{gao2022frac}. We make \param{four} key observations from \figref{fig:maj3_success}. First, \X{} achieves \param{97.91}\% \yct{(up to \param{100\%})} success rate by activating thirty-two rows, \param{24.18}\% higher success rate than the FracDRAM on average. Second, the data pattern significantly affects the success rate of \maj{3} operation. We hypothesize that this occurs due to interference between cells located in close proximity, as demonstrated in prior work~\cite{khan2016parbor}. Therefore, this phenomenon affects the deviation on a bitline during charge-sharing, leading to incorrect results. Third, in all module groups, increasing N results in a higher success rate as it makes the deviation on the bitline closer to the safe margins, as explained in \secref{subsec:inp_repl}. Fourth, Mfr. M has a higher success rate than Mfr. H in all \nrg{}. We hypothesize that Mfr. M can have more robust sense amplifiers than Mfr. H. This can allow the sense amplifier to safely amplify the reduced deviation on the bitline voltage correctly. We conclude that input replication greatly increases the success rate of \maj{3} operation in all tested modules.

\begin{figure}[ht]
\centering
\includegraphics[width=0.85\linewidth]{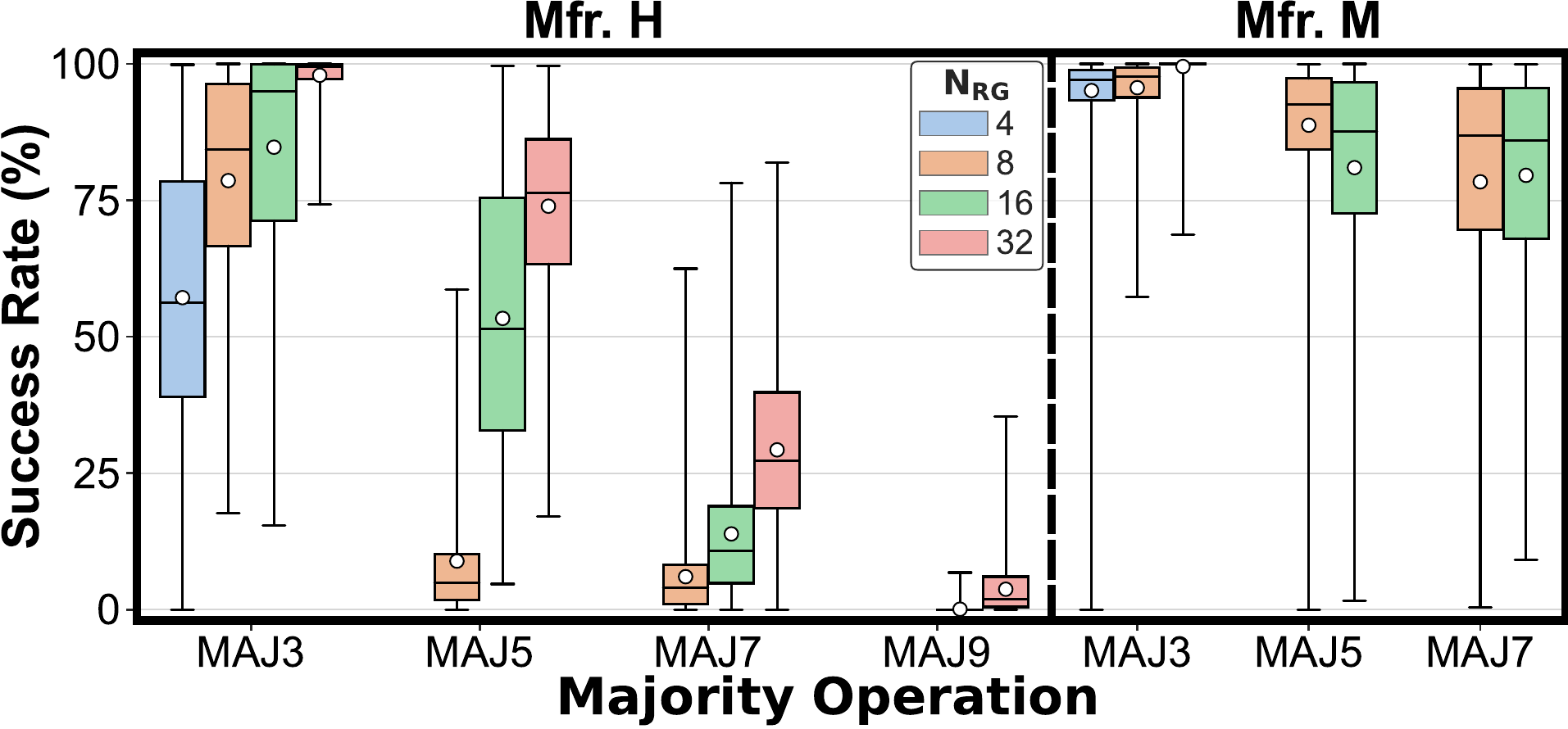}
\caption{\maj{3}, \maj{5}, \maj{7}, and \maj{9} success rate of \nrg{} for every N value across different DRAM manufacturers.}
\label{fig:maj579_success}
\end{figure}

\figref{fig:maj579_success} shows a box-and-whiskers plot\footref{fn:boxplot} of the \maj{3}, \maj{5}, \maj{7}, and \maj{9} success rate of \nrg{} for every N value (x-axis) across all \nMODULES{} modules we test using random data pattern. 
We omit the majority operations (i.e., MAJ11+ for Mfr. H and MAJ9+ for Mfr. M) that have $<$1\% success rate on average. We make \param{three} key observations from \figref{fig:maj579_success}. First, \X{} can reliably perform \maj{5} operation with \param{73.93}\% \yct{(up to \param{99.61}\%)} and \maj{7} operation with \param{29.28}\% (up to \param{81.92}\%) success rate. Second, as the number of inputs of the majority operation increases, the success rate decreases. We hypothesize that when we increase the number of inputs, the number of copies from each input decreases, making the deviation on the bitline closer to unreliable sensing margins. Third, Mfr. M outperforms Mfr. H in every \maj{M} in terms of success rate, which can be due to the hypothesis in \figref{fig:maj3_success}'s observations.
We conclude that by leveraging input replication, \X{} increases the success rate of the majority operation regardless of the number of input operands in both manufacturers.

\begin{figure}[ht]
\centering
\includegraphics[width=0.8\linewidth]{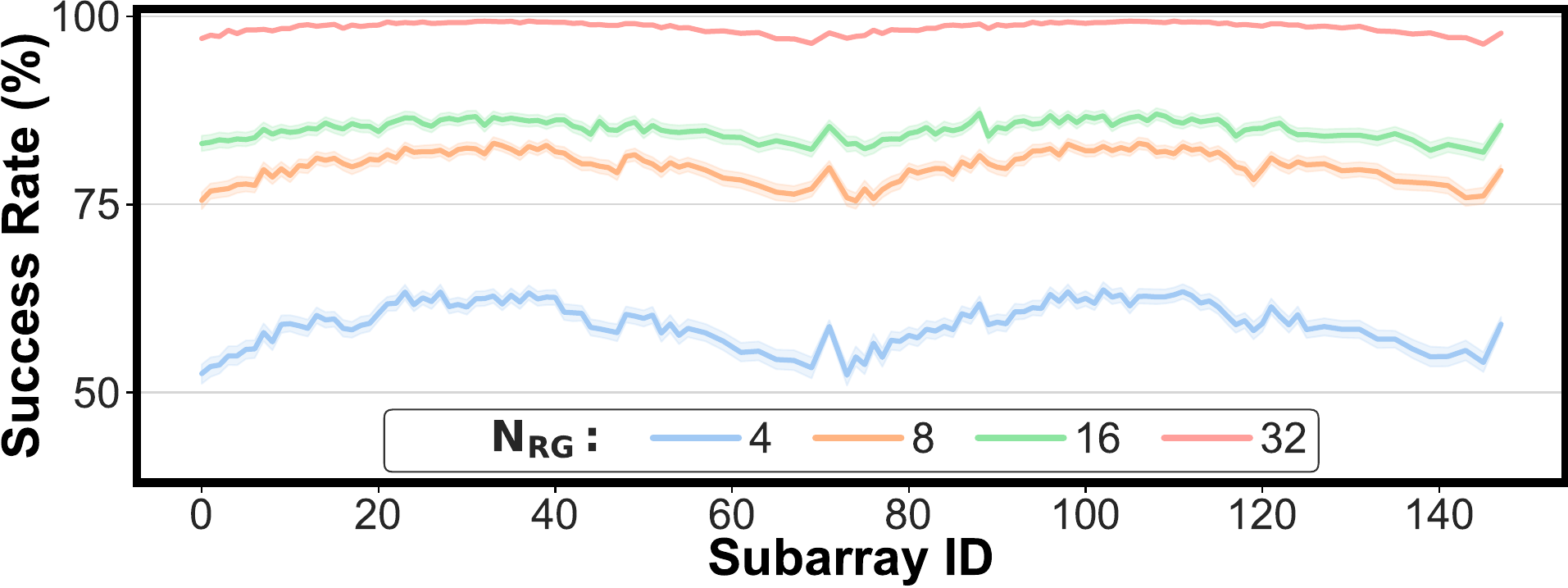}
\caption{Average \maj{3} success rate across all subarrays in a DRAM bank.}
\label{fig:maj3_variation}
\end{figure}

\noindent
\textbf{Spatial Distribution of Success Rate.} We study the spatial distribution of success rate of \maj{3} across every subarray in a DRAM bank of \param{H0} module. In each subarray, we randomly select 100 \nrg{} for every N and perform \maj{3} operation using a random data pattern. \figref{fig:maj3_variation} depicts how a \nrg{}'s average success rate varies across subarrays in a DRAM bank.

We make two key observations from \figref{fig:maj3_variation}. First, \X{} significantly increases the success rate of the majority operation in every subarray on average. Second, the overall success rate distribution follows an M-like pattern. The success rate peaks in the first quarter of subarrays and descends in the second quarter of subarrays. This trend repeats itself in the second half of the \yct{bank}. We hypothesize that this pattern results from the effects of systematic process variation. We conclude that regardless of the spatial location of an \nrg{}, \X{} outperforms the FracDRAM by \param{66.23\%} in average success rate across all subarrays.

\subsubsection{Majority-based Computation}
In this section, we study the potential benefits of enabling \maj{M} operations in off-the-shelf DRAM chips on microbenchmarks. We analyze 1) performance gain (i.e., speedup on execution time) using new majority operations and 2) the sensitivity of a number of rows that are simultaneously activated (\nrg{}) to performance gain.

Majority operation can be used to implement 1) logic operations such as AND/OR~\cite{hajinazar2020simdram,ali2019memory,seshadri.arxiv16,seshadri2017ambit,seshadri2015fast,seshadri2019dram,gao2019computedram}) and XOR operations~\cite{alkaldy2014novel}, and 2) full adder operations~\cite{ali2019memory,gao2019computedram,hajinazar2020simdram}. These operations are then used as basic building blocks for the target in-DRAM computation (e.g., addition, multiplication)~\cite{ali2019memory,angizi2019graphide,li2016pinatubo,gao2019computedram}.

\noindent
\textbf{Real DRAM Chip Experiments.} We tightly schedule the DRAM commands to perform majority operations and measure the execution time using the DRAM Bender. We evaluate the execution time of seven arithmetic \& logic microbenchmarks for two vendors (\maj{3}, \maj{5}, and \maj{7} for Mfr. M and \maj{3}, \maj{5}, \maj{7}, and \maj{9} for Mfr. H). For each majority operation, we choose the \nrg{} that produces the highest throughput across all \nMODULES{} tested DRAM modules. We perform 32-bit logic (bitwise and, or, and xor reductions) and arithmetic (addition, subtraction, multiplication, and division) computations on 8KB elements. We use 65536-element (DRAM row size) two-input vectors (e.g., A and B) where each element of the vectors is a 32-bit integer. Each element of A and B that has the same index (e.g., A[X] and B[X] in column X) is stored in the same column.

We evaluate PULSAR by employing the framework described in the prior work~\cite{gao2019computedram}, which is based on bit-serial computation and stores the negated value of operands in the same subarray as the original operands, computed beforehand in the CPU. Bit-serial computation, using a vertical layout where operands are aligned along bitlines, applies bulk bitwise operations to entire rows of DRAM, generating results from bitlines in parallel. This approach enables PuM to perform operations efficiently~\cite{hajinazar2020simdram,ali2019memory,gao2019computedram,seshadri2017ambit}. We refer the reader to the prior work~\cite{gao2019computedram} for the details of the framework.

\begin{figure}[ht]
\centering
\includegraphics[width=0.85\linewidth]{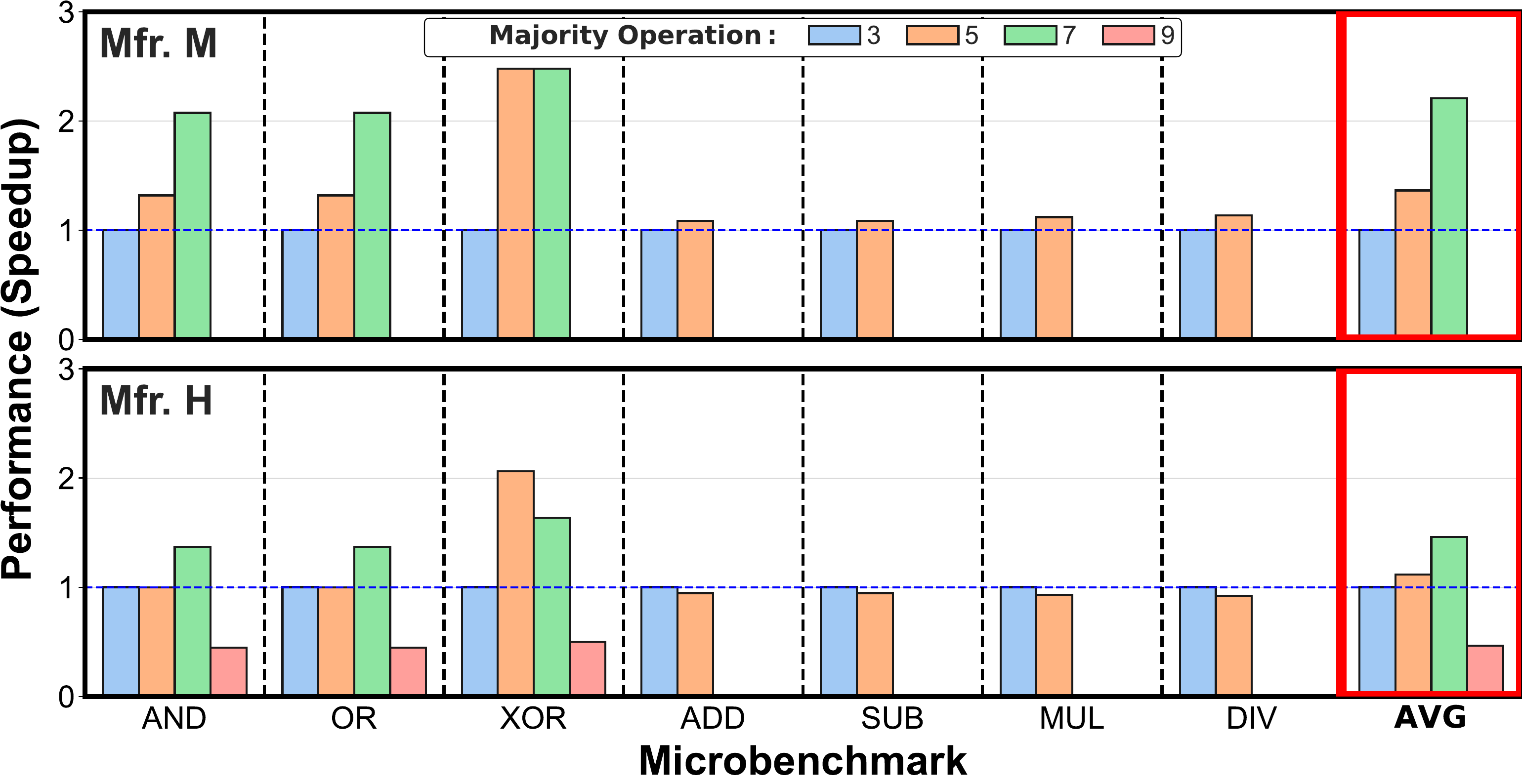}
\caption{Speedup over the state-of-the-art (\maj{3}) in \yct{seven} arithmetic \& logic microbenchmarks.}
\label{fig:maj_perf_1}
\end{figure}

\figref{fig:maj_perf_1} shows the performance of the majority operations of two manufacturers in seven microbenchmarks normalized to the state-of-the-art mechanism, FracDRAM~\cite{gao2022frac} (i.e., \maj{3} with \param{four} rows), which is the blue dashed line. We make three key observations. First, \X{} outperforms FracDRAM in all microbenchmarks. On average, \X{} provide \param{2.21}$\times$ (\param{1.46}$\times$) performance improvement over FracDRAM in Mfr. M (Mfr. H). Second, increasing the number of operands in the \maj{} provides more performance. \maj{7} provides \param{1.62}$\times$ (\param{1.31}$\times$) of the performance improvement provided by \maj{5} in Mfr. M (Mfr. H). Third, in Mfr. H, \maj{9} incurs performance degradation by \param{2.14}$\times$. This is because \maj{9} has a poor success rate (maximum \param{35.35\%} success rate, shown in \figref{fig:maj579_success}), which requires repeatedly performing the \maj{9}, resulting in higher latency. We conclude that \X{} significantly achieves \param{2.21}$\times$ better performance than FracDRAM by enabling new \pum{} primitives.

\noindent
\textbf{Sensitivity to \nrg{}.} We study the effect of the number of rows that are activated simultaneously \nb{(\nrg{})} on majority-based computation performance. \nb{Increasing \nrg{} increases the success rate due to input replication. However, it can also increase the initialization cost since more rows are required to be initialized.} We evaluate the execution time of seven microbenchmarks based on majority operations. To further analyze the potential benefits and limitations, we study the impact of initialization latency and the success rate on the performance of microbenchmarks for various numbers of rows that are used to realize the majority operation. \nb{We study four different scenarios: 1)  RealExp: real experiment results, i.e., using empirical latency and success rate values, 2) RealInit: 100\% success rate with empirical latency, 3) RealSR: empirical success rate with no initialization latency, and 4) Ideal: 100\% success rate with no initialization latency. }

\begin{figure}[ht]
\centering
\includegraphics[width=0.85\linewidth]{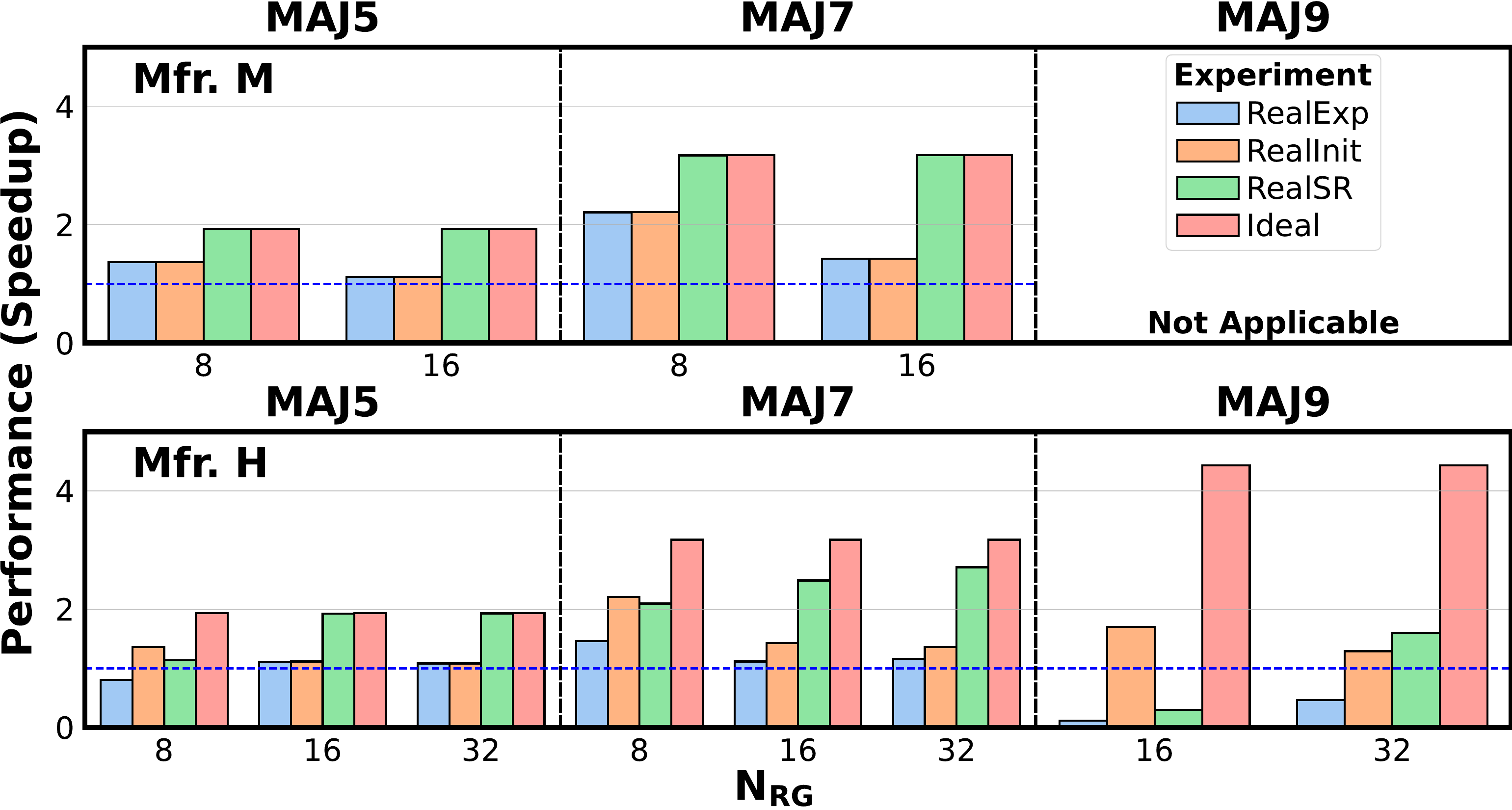}
\caption{Performance sensitivity to \nrg{} of Mfr. M (top) and Mfr. H (bottom) modules. All bars represent the average speedup over FracDRAM across seven microbenchmarks.}
\label{fig:maj_sens}
\end{figure}

\figref{fig:maj_sens} shows the \nb{average} speedup of using 8, 16, and 32 rows to perform \maj{M} over the FracDRAM \nb{across all microbenchmarks}. We make two key observations from \figref{fig:maj_sens} for Mfr. M. First, \nb{increasing the success rate results in only negligible performance improvement due to the already high empirical success rate (\param{100}\% for \maj{5} and \param{99.95}\% for \maj{7})}. Second, in all \maj{M}, since the success rate is high, increasing the number of rows only increases the overhead of initialization latency and thus degrades the performance. \nb{We make two key observations for Mfr. H. First,} providing a 100\% success rate increases the performance by \param{2.55}$\times$ on average as Mfr. H has low empirical success rate (\param{99.61}\% for \maj{5} and \param{81.92}\% for \maj{7}, and \param{35.35}\% for \maj{9}). Second, increasing the number of rows can improve the performance as it enables a better success rate. We conclude that 
for both Mfr. H and Mfr. M, reducing the initialization latency improves the performance of \maj{M} operations that have a high success rate and can even achieve maximum performance. 

%% file: sections/06-3_bulk-init.tex
\subsection{Content Destruction for Cold Boot Attack}
\label{subsec:coldboot}

A cold boot attack is a physical attack on DRAM that involves hot-swapping a DRAM chip and reading out the contents of the DRAM chip ~\cite{bauer2016lest,gruhn2013practicality,halderman2009lest,hilgers2014post,lee2012correcting,lindenlauf2015cold,muller2010aesse,simmons2011security,villanueva2019cold,yitbarek2017cold}.
Cold boot attacks are possible because the data stored in DRAM is not immediately lost when the chip is powered off. This is due to the capacitive nature of DRAM cells that can hold their data up to several seconds~\cite{bauer2016lest,khan2014efficacy,liu2013experimental,liu2012raidr,patel2017reaper} or minutes~\cite{halderman2009lest}.
This effect can be exacerbated with low temperatures, resulting in DRAM cells retaining their content even longer.

A practical and secure way to mitigate cold boot attacks is to destroy the DRAM content rapidly during power-off/on~\cite{orosa2021codic,tcg2008platform}. \X{} can quickly write a predetermined value (e.g., all-zeros) to many rows with \emph{Bulk-Write} and copy this value to many other rows \agy{using} \mrc{}. This way, \X{} can be used to \emph{rapidly} destroy the DRAM content.

\noindent
\textbf{Evaluation.} 
We evaluate \X{}-based content destruction with varying numbers of rows that are simultaneously activated, from 2 to 32. \iey{\X{}-based content destruction with N-row activation can leverage up to N-row activation (e.g., 16-row activation can use 2-, 4-, 8-, and 16-row activation but cannot use 32-row activation). \X{}-based content destruction choose the \nrg{}s with a greedy algorithm to effectively destruct the contents of all rows in a bank by issuing the least number of \apa{} command sequence.}
We compare \X{}-based content destruction to \one{} RowClone~\cite{gao2019computedram}-based and \two{} Frac\agy{DRAM}~\cite{gao2022frac}-based content destruction. 
The RowClone-based content destruction is implemented as a two-step process. First, it issues a \wri{} command to write predetermined data to an arbitrary row. Second, it performs RowClone to overwrite the content of the DRAM rows, making the original content inaccessible. 
The Frac\agy{DRAM}-based content destruction is implemented to repeatedly send the Frac operation to every row to put the rows into a neutral state, making them store \vddh{}. We schedule the DRAM commands to perform all content destruction operations (i.e., Bulk-Write, \mrc{}, RowClone, and Frac) and measure the execution time to overwrite all the data in a bank of off-the-shelf DRAM module (H7).

\begin{figure}[ht]
\centering
\includegraphics[width=0.78\linewidth]{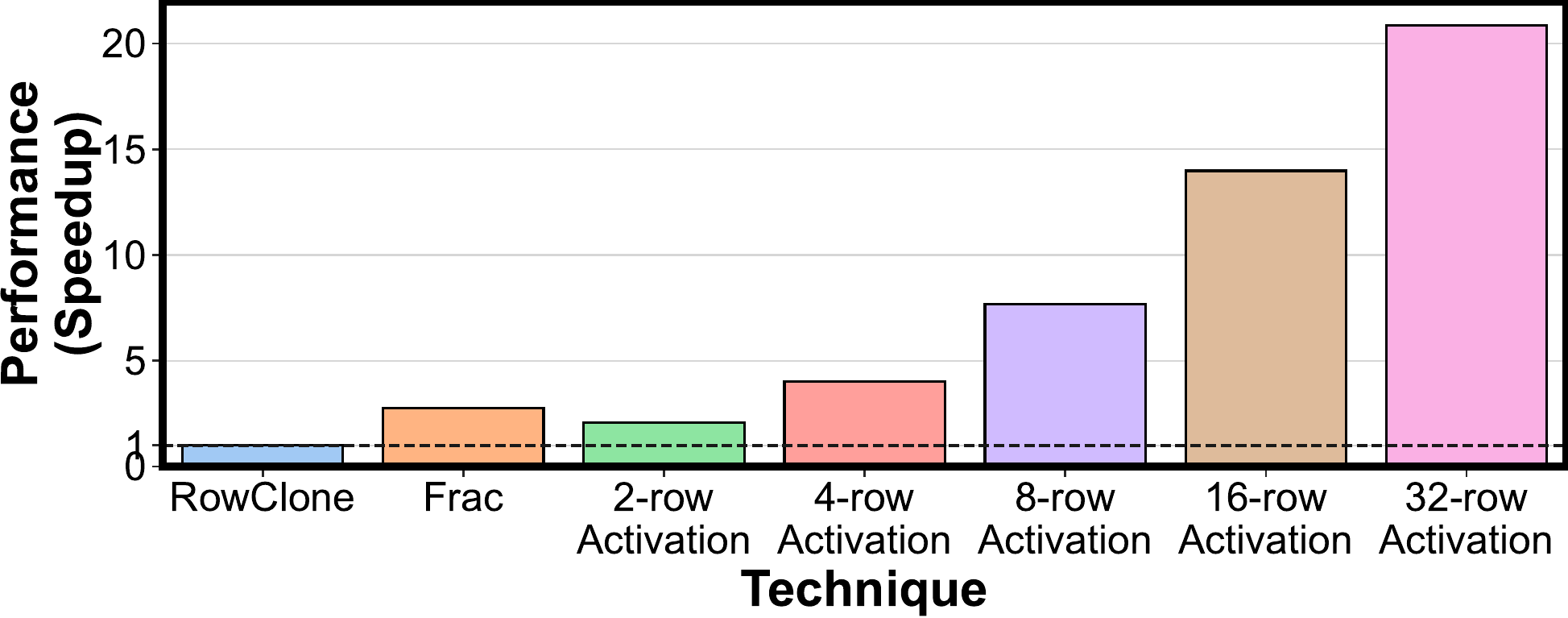}
\caption{Speedup over the RowClone-based content destruction in a DRAM bank.}
\label{fig:cold_boot}
\end{figure}

\figref{fig:cold_boot} \agy{shows} the speedup in execution time for content destruction normalized to the RowClone-based content destruction's execution time. 

\nb{We make two key observations \agy{based on the \figref{fig:cold_boot}}. 
First, \X{}-based content destruction with 4-, 8-, 16-, 32-row activation outperforms both RowClone-based content destruction and \agy{FracDRAM}-based content destruction up to \param{20.87}$\times$ and \param{7.55}$\times$, respectively. 
Second, increasing the number of simultaneously activated rows increases the speedup of \X{}\-/based technique.
Because increasing the number of operands in \mrc{} and Bulk-Write decreases the total number of operations.}
We conclude that \X{}\nb{-based content destruction outperforms both techniques and destroys \agy{DRAM} content significantly faster than the state-of-the-art techniques.}

%% file: sections/07_discussion.tex
\section{Discussion}

\agy{This section} discuss\agy{es} \one{} how \X{} can be leveraged in addition to \agy{the} use cases in \secref{sec:use-cases}, and \two{} \agy{\X{}'s limitations}.

\noindent
\textbf{Extending \X{}\agy{'s Use Cases.}}
\label{subsec:extending}
\agy{We demonstrate the effectiveness of \X{} in two use cases. However, \X{} can be leveraged to improve other \pum{} systems. This section discusses two additional use cases.}
\agy{First, an end-to-end framework, using a DRAM chip as a \pum{} substrate, can leverage \X{}.}
\agy{For example,} SIMDRAM~\cite{hajinazar2020simdram}
\agy{automatically creates} desired complex operations (e.g., addition and multiplication)
\agy{by employing} majority-inverter graphs to accelerate a broad range of workloads, including graph processing, databases, neural networks, and genome analysis. Unfortunately, SIMDRAM \agy{uses} \emph{only} \maj{3} \agy{due to the low success rate of majority operations with more inputs} (e.g., \maj{5}). 
\agy{SIMDRAM can be extended by leveraging \X{}'s input replication technique to reliably execute \maj{} operations with more inputs (e.g., \maj{5}, \maj{7}, and \maj{9}) and thus, to achieve higher performance.} However, an end-to-end system needs to address several key challenges, such as (1) programming interface, (2) compiler support, and (3) end-to-end system integration. Designing an end-to-end system for \X{} is a direction that future work can explore.
\agy{Second, \X{} can be used to generate \gls{puf} and \gls{trn}. Prior works experimentally demonstrate that it is possible to generate high throughput \gls{puf}~\cite{kim2018dram} and \gls{trn}~\cite{kim2019drange,olgun2021quactrng} in off-the-shelf DRAM chips by violating timing constraints. Unfortunately, the throughputs of these works are bound by the latency of initializing multiple DRAM rows before each \gls{puf} and \gls{trn} generation. These works can use \X{}'s \mrc{} and bulk-write primitives to reduce their initialization latency. We leave the exploration of these use cases for future work.}

\noindent
\textbf{Limitations.}
We identify limitations of \X{} under four categories.
First, \agy{even though off-the-shelf DRAM chips allow simultaneously activating multiple DRAM rows, they do \emph{not} provide the user with the flexibility of choosing which rows to activate.}
Second, all \agy{of the tested} DRAM chips \agy{that successfully perform} multiple-row activation is from Micron \agy{(Mfr. M)} and SK Hynix \agy{(Mfr. H)}. We conduct experiments on 64 DRAM chips from one major manufacturer, Samsung. Unfortunately, \agy{we do \emph{not} observe a successful} multiple-row activation \agy{in any of the tested} Samsung chips. We hypothesize that these DRAM chips have internal circuitry that ignores the \pre{} command or the second \act{} command when the timing parameters (\gls{trp} and \gls{tras}) are greatly violated\agy{, which agrees with the hypotheses of prior work~\cite{yauglikcci2022hira}}.\agy{Unlike prior works~\cite{gao2022frac,gao2019computedram,olgun2021quactrng,yauglikcci2022hira}, \X{} achieves a high success rate on DRAM chips from Mfr. M, which requires a deep understanding of the hierarchical row decoder to choose the set of DRAM rows to target for two \act{} commands.}
Third, \X{} is capable of performing many-input majority operations (theoretically, up to \maj{31}). However, \X{} cannot reliably perform majority operations with more than nine inputs (i.e., \maj{9+}) \agy{due to very low success rates~(\secref{subsubsec:maj_success}).} Four, \X{} potentially have an effect on transient errors in DRAM chips. In our experiments, described in \secref{subsubsec:maj_success}, we check for bitflips in the whole DRAM bank. We do not observe any errors in rows outside of the row group across any of the tested DRAM chips. We believe that investigating all potential effects of PULSAR on any type of transient error requires rigorous analysis and extensive exploration, which warrants its own study.

\X{} is not an execution model that is immediately usable. We demonstrate a proof-of-concept of performing multi-row activation in real off-the-shelf DRAM chips and its potential benefits in improving the success rate and the performance of previously proposed PuM operations. Our work contributes towards a better understanding of the capability of real off-the-shelf DRAM chips. We hope and expect that DRAM manufacturers will adopt our approach in future DRAM chips and officially support \X{}. We conclude that none of these limitations fundamentally prevent a system designer from using off-the-shelf DRAM chips to perform \pum{} operations and thus benefit from \X{}'s high reliability and performance \agy{benefits}. \agy{We hope and expect future DRAM chips to officially support simultaneous many-row activation and alleviate all of these limitations.}

%% file: sections/08_related-work.tex
\section{Related Work}
\label{sec:related-work}

\nb{To our knowledge, this is the first work that demonstrates a proof-of-concept that off-the-shelf DDR4 DRAM chips are capable of simultaneously activating up to 32 rows. \X{} leverages this new observation and improves the success rate and the performance of \pum{} operations compared to the state-of-the-art \pum{} technique~\cite{gao2022frac}.}

\noindent
\textbf{Multiple Row Activation in Off-the-shelf DRAM.}
\nb{Many prior works propose various forms of \pum{} operations in off-the-shelf DRAM devices using multiple row activation~\cite{olgun2021quactrng,gao2019computedram,gao2022frac,yauglikcci2022hira}. ComputeDRAM~\cite{gao2019computedram} presents a DRAM command sequence (\apa{}) enabling the triple row activation, resulting in a bitwise AND/OR function by violating timing parameters between consecutive DRAM commands. FracDRAM\cite{gao2022frac} stores fractional values in off-the-shelf DDR3 devices by employing a DRAM command sequence (\act{} $\rightarrow$ \pre{}) with reduced timing parameters. By leveraging the fractional values stored in DRAM, FracDRAM provides an improved success rate in \maj{3} operation and implements a physical unclonable function in DRAM.} FracDRAM observes that up to 16 rows can be simultaneously activated in off-the-shelf DDR3 chips. However, FracDRAM does not provide any characterization or hypothesis of the reason behind this observation. \X{} introduces many row activations (up to 32 rows) by choosing the target rows that are activated carefully. \X{} proposes a hypothetical row decoder design that explains how many rows can be activated simultaneously. \nb{\X{} improves the success rate of existing \maj{3} operations and improves the performance of \pum{} applications by introducing new \pum{} primitives based on many row activation. }

\nb{Other works enable different functionalities using simultaneous many-row activation. QUAC-TRNG~\cite{olgun2021quactrng} introduces quadruple row activation and exploits this phenomenon to generate true random numbers in off-the-shelf DRAM chips. QUAC-TRNG proposes a hypothetical row decoder design that enables quadruple row activation. HiRA~\cite{yauglikcci2022hira} introduces a hidden row activation mechanism by simultaneously opening two rows, leveraging the hidden row activation to implement a refresh-based RowHammer mitigation mechanism. \X{} can be used to potentially extend these mechanisms as these proposals leverage many-row activation.}

\noindent
\textbf{Other Off-the-shelf-DRAM-based PuM.}
Prior works design off-the-shelf-DRAM-based mechanisms to implement TRNG and PUF. DRAM-based TRNGs generate true random numbers by violating timing parameters~\cite{kim2019drange,talukder2019exploiting}, using retention-based failures~\cite{keller2014dynamic,sutar2016d} and using startup values~\cite{eckert2017drng,tehranipoor2016robust}. DRAM-based PUFs generate device-specific signatures using retention-based failures~\cite{keller2014dynamic,sutar2016d,xiong2016run}, by violating timing parameters~\cite{kim2018dram}, by exploiting write access latencies~\cite{hashemian2015robust}, and using startup values~\cite{tehranipoor2016dram}. These operations can leverage the functionality of \X{} to reduce their initialization latency, thereby increasing their throughput.

\noindent
\textbf{Modified-DRAM-based PuM.} Prior works propose modification into DRAM to perform \pum{} operations~\cite{li2016pinatubo,li2017drisa,Li2018SCOPEAS,manning2018apparatuses,oliveira2022accelerating,parveen2017hybrid,parveen2017low,parveen2018hielm,parveen2018imcs2,rakin2018pim,ramanathan2020look,rezaei2020nom,seshadri.arxiv16,seshadri2013rowclone,seshadri2015fast,seshadri2015gather,seshadri2016processing,seshadri2017ambit,seshadri2017simple,seshadri2018rowclone,seshadri2019dram,shafiee2016isaac,song2017pipelayer,song2018graphr,tian2017approxlut,wu2022dram,xie2015fast,xin2019roc,xin2020elp2im,yang2020flexible,yu2018memristive,zawodny2018apparatuses,zha2020hyper,zhao2017apparatuses}. RowClone~\cite{seshadri2013rowclone} is a low-cost DRAM architecture that can perform bulk data movement operations inside DRAM chips. Ambit~\cite{seshadri2017ambit} modifies the DRAM circuitry to perform bitwise \maj{3} (and thus bitwise AND/OR) by activating three rows simultaneously and bitwise NOT operations in DRAM. Unfortunately, these mechanisms require changes to DRAM chips and are not applicable to off-the-shelf DRAM chips.

%% file: sections/09_conclusion.tex
\section{Conclusion}

We introduce \X{}, a proof-of-concept technique that enables high-success-rate and high-performance \gls{pum} operations in off-the-shelf DRAM chips. \X{} leverages the key observation that by issuing a carefully crafted sequence of DRAM commands, up to 32 rows can be activated simultaneously. \X{} improves 1) the success \yct{rate} through input data replication and 2) performance by enabling new \pum{} primitives. Our experimental results, conducted on \nCHIPS{} off-the-shelf DDR4 DRAM chips from two major manufacturers, demonstrate the effectiveness of \X{} on two use cases. \X{} achieves significant improvement over the state-of-the-art \yct{in terms of} both success rate and performance. Our results show that compared to the state-of-the-art mechanism, \X{} has 24.18\% higher success rate and improves the performance in majority-based microbenchmarks by $2.21\times$ on average. 

\section*{Acknowledgements}
We thank the SAFARI Research Group members for providing a stimulating intellectual environment. 
We acknowledge the generous gifts from our industrial partners, including Google, Huawei, Intel, and Microsoft. 
This work is supported in part by the Semiconductor Research Corporation (SRC), the ETH Future Computing Laboratory (EFCL), 
and the AI Chip Center for Emerging Smart Systems (ACCESS).

%% file: sections/appendix.tex
\appendix
\section{\X{} System Integration}

\noindent\textbf{Power Constraints.}~We count \X{}'s row activations and issue them with respect to the~\gls{tfaw} constraint in DDRx DRAM standards~\cite{jedec2017ddr4,jedec2020ddr5,jedec2015lpddr4,jedec2020lpddr5}, which limits the rate of performed activations in a rank to stay under the power budget.~Hence, we ensure that the row activations are performed within the power budget of a DRAM rank.

\noindent\textbf{Compatibility with Off-the-{Shelf} DRAM Chips.}
We experimentally demonstrate that \X{} works on \nCHIPS{} off-the-shelf DDR4 DRAM chips from two major DRAM manufacturers. Therefore, \X{} does \emph{not} require any modifications to these {real} DRAM chips.

\noindent\textbf{Compatibility with {Different Computing} System{s}.}
We discuss \X{}'s compatibility with three {types} of computing systems:~1) FPGA-based systems {(e.g., PiDRAM \cite{olgun2021pidram, olgun2022drambender})}, {2) c}ontemporary processors, and {3) s}ystems with programmable memory controllers~{\cite{tassadaq2014pmss, bojnordi2012pardis}}.
First, \X{} can be easily integrated into all existing FPGA-based systems that use DRAM to store data~\cite{olgun2021pidram, xilinx-data-center, xilinx-fpga, olgun2022drambender}. \agy{We showcase a system integration using DRAM Bender~\cite{olgun2022drambender} for our performance evaluation as it is widely available and does \emph{not} require any changes in the processor circuitry (\secref{sec:end-to-end}).}
Second, contemporary processors require modifications to their memory controller logic to implement \X{}. Implementing \X{} is a design-time decision that requires balancing manufacturing cost with \X{}'s performance benefits. 
{We show that \X{} significantly improves system performance (\secref{sec:end-to-end}), but \nb{we} leave the analysis of such integration's hardware complexity for future work.}
Third, systems that employ programmable controllers~\cite{tassadaq2014pmss,bojnordi2012pardis} can be {relatively} easily modified to implement \X{} {by} programming \X{}'s operations using the ISA of programmable memory controllers~\cite{tassadaq2014pmss,bojnordi2012pardis}.

\section{Effect on Real-World Kernels}
\label{sec:end-to-end}
\agy{W}e present a comprehensive evaluation to provide insights into \X{}'s performance benefits on nine real-world kernels \atb{over a real CPU, a GPU} and state-of-the-art commodity DRAM-based PuM techniques.
\subsection{Experimental Methodology}
\noindent\textbf{Experimental Setup.} We evaluate \X{} using a \emph{real} end-to-end system that consists of two components: \one{} \atb{a contemporary computer that hosts the workloads we evaluate (host machine)} and \two{} a\atb{n FPGA board that implements} DRAMBender~\cite{olgun2022drambender} connected to the host \atb{machine} through a PCIe bus. We extend the DRAMBender C++ API to support tightly scheduling DRAM commands for performing \X{} (i.e., \mrc{}, MAJ3, MAJ5, and MAJ7) and FracDRAM (i.e., RowClone and MAJ3) operations.

\noindent\textbf{Algorithm for evaluating \X{}/FracDRAM.} We evaluate \X{} and FracDRAM in \param{three} steps:
First, the host \atb{machine} computes the input operands' negated values, and both the original and negated data are then stored in the FPGA board's DRAM module in a vertical data layout (\secref{back:maj}). 
Second, we create a DRAM Bender program that \atb{implements the workload we test using PULSAR's new computation primitives (e.g., MAJ5)}, and we offload the program to the FPGA \nb{b}oard to perform PuM operations. 
Third, \atb{the} DRAM \atb{module} performs PuM operations, and the results of the PuM operations are read back from the DRAM module to the application running on the host machine. 
We repeat this process for each workload, capturing the execution time of each workload \iey{by taking the PCIe latency into account.}

\subsection{Results}
We analyze the performance benefits of \X{} on \agy{real-world applications} and compare \X{} against CPU, GPU, and FracDRAM. \agy{We} use a real multicore CPU (Intel Skylake~\cite{i7-6700K}) \agy{and optimize our workloads} to leverage AVX-512 instructions~\cite{firasta2008intel}. \agy{We measure performance on} a real high-end GPU (Nvidia Titan V~\cite{TitanX}). We capture GPU kernel execution time that excludes data initialization/transfer time. We report the average of five runs for each CPU/GPU data point, each with a warmup phase, to avoid cold cache effects. We capture the execution time of each workload on CPU and GPU.

We conduct evaluations on \agy{9 real-world applications that heavily rely on the evaluated microbenchmarks.~We explain these applications under five categories.}~\textit{\one{} Convolutional Neural Networks (CNNs)}: We use XNOR-NET implementations~\cite{rastegari2016xnor} of VGG-13, VGG-16, and LeNet-5 provided by~\cite{he2020sparse}, which performs convolutions using a series of bitcount, addition, and XNOR operations. We evaluate the inferences of \agy{VGG-13 and -16 using CIFAR-10~\cite{krizhevsky2010convolutional} and LeNeT-5 using MNIST~\cite{deng2012mnist} datasets.}~\textit{\two{} k-Nearest Neighbor Classifier (kNN)}: We apply the kNN classifier to solve the handwritten digits recognition problem using the MNIST dataset. \agy{We implement a quantized 8-bit version of the} Euclidean distance algorithm entirely in DRAM using \X{}. 
\noindent\textit{\three{} Database}: We evaluate two workloads\agy{: BitMap Indices (BMI)~\cite{chan1998bitmap} and BitWeawing (BW)~\cite{li2013bitweaving}. BMI provides} high space efficiency and high performance for many queries (e.g., join and scan) in databases compared to traditional B-tree indices. \agy{Our BMI workload}
runs the query: \agy{``}How many users were active every day for the past month?\agy{''} \agy{on} a database that tracks the login activities of 8 million users.
\agy{Our BW workload evaluates} a simple table scan query\agy{:} \texttt{select count(*) from T where c1 <= val <= c2}\agy{.}~\textit{\four{} Graph Processing}: We evaluate two graph procesing workloads\agy{: k-Clique Star (KCS)~\cite{besta2021sisa,besta2021graphminesuite} and Triangle Counting (TC)~\cite{besta2021sisa,besta2021graphminesuite}}.
KCS aims to find all k-clique stars in a given graph. A k-clique star consists of a k-clique (a set of k fully connected vertices) and additional vertices connected to all k-clique members. Using a set-centric approach~\cite{besta2021sisa}, we represent vertices \agy{and k-cliques in the form of} bit-vectors encoding their adjacency to others, which enables us to perform this operation by a set of bitwise operations, similar to~\cite{flashcosmos,besta2021sisa}. 
TC involves calculating the total number of 3-cycles (triangles) in a graph, and it can be done by a set of bitwise operations, similar to~\cite{besta2021sisa}.
\textit{5) Image processing}: image segmentation (IMS), an image processing application that aims to break an image into multiple regions depending on a given set of colors. In IMS, each image consists of 800×600 pixels with four colors. We adapt our implementation using the prior PuM works~\cite{flashcosmos,gao2021parabit}.

We evaluate two different configurations of \X{} and FracDRAM where 1 (\X{}:1 and FracDRAM:1) and 16 (\X{}:16 and FracDRAM:16) banks out of all the banks in the DRAM module to leverage bank-level parallelism to maximize DRAM throughput~\cite{kim2010thread,salp,Kim2016Ramulator,lee2009improving,mutlu2008parallelism}.

\begin{figure}[ht]
\centering
\includegraphics[width=\linewidth]{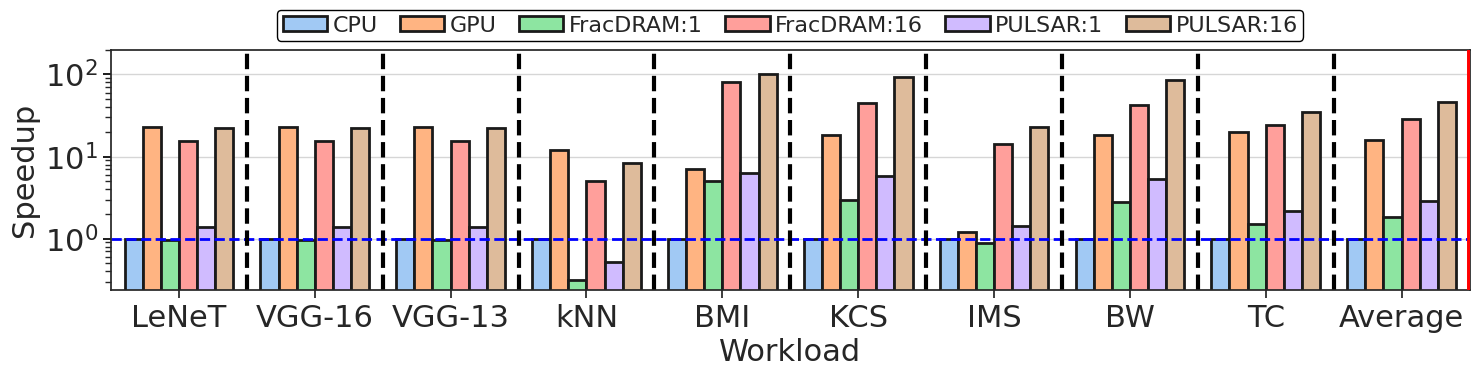}
\caption{Normalized speedup of real-world applications. \X{}:X and FracDRAM:X uses X DRAM banks for computation.}
\label{fig:real_world}
\end{figure}

\figref{fig:real_world} shows the performance of \X{} and our baseline configurations for each application, normalized to that of the multicore CPU. We make three key observations. First, \X{}:16 greatly outperforms the CPU and GPU baselines, providing \param{43.38$\times$} and \param{2.65$\times$} the performance of the CPU, and GPU, respectively, on average across all nine applications. Second, \X{}:16 (\X{}:1) provides \param{1.59$\times$} (1.55$\times$) the performance of FracDRAM:16 (FracDRAM:1), on average, across all nine applications, with a maximum of \param{2.01$\times$} (\param{1.90$\times$}) the performance of FracDRAM:16 (FracDRAM:1) for the BW application. Third, even with a single DRAM bank (i.e., \X{}:1), \X{} always outperforms the CPU baseline, providing \param{2.71$\times$} the performance of the CPU on average across all applications. This speedup is a direct result of leveraging the high in-DRAM bandwidth in \X{} to avoid the memory bottleneck in the CPU caused by the large amounts of intermediate data generated in such applications. We conclude that \X{} is an effective and efficient off-the-shelf DRAM-based technique to accelerate many commonly-used real-world applications.